\documentclass[aps,pra,twocolumn,superscriptaddress]{revtex4-2}
\usepackage{graphicx}
\usepackage{amsmath}
\usepackage{amsbsy}
\usepackage{array}
\usepackage{amssymb}
\usepackage{verbatim}
\usepackage{caption} 
\usepackage{subcaption} 
\usepackage{float} 
\usepackage{hyperref}
\usepackage{epstopdf}
\hypersetup{
	colorlinks=true,  
	linkcolor=blue,   
	citecolor=blue,   
	urlcolor=blue,    
	pdfborder={0 0 0}, 
}

\begin{document}
\title{Characterizing the set of quantum correlations in prepare-and-measure quantum multi-chain-shaped networks}

\author{Yanning Jia}
\affiliation{School of Science, Beijing University of Posts and Telecommunications, Beijing 100876, China}
\affiliation{Key Laboratory of Mathematics and Information Networks, Beijing University of Posts and Telecommunications, Ministry of Education, China}
\affiliation{State Key Laboratory of Networking and Switching Technology, Beijing University of Posts and Telecommunications, Beijing, 100876, China}

\author{Fenzhuo Guo}\email{gfenzhuo@bupt.edu.cn}
\affiliation{School of Science, Beijing University of Posts and Telecommunications, Beijing 100876, China}
\affiliation{Key Laboratory of Mathematics and Information Networks, Beijing University of Posts and Telecommunications, Ministry of Education, China}
\affiliation{State Key Laboratory of Networking and Switching Technology, Beijing University of Posts and Telecommunications, Beijing, 100876, China}
\author{YuKun Wang}
\affiliation{Beijing Key Laboratory of Petroleum Data Mining, China University of Petroleum-Beijing, Beijing, 102249, China.}
\author{Haifeng Dong}
\affiliation{School of Instrumentation Science and Opto-Electronics Engineering, Beihang University, Beijing,100191,China}

\author{Fei Gao}
\affiliation{State Key Laboratory of Networking and Switching Technology, Beijing University of Posts and Telecommunications, Beijing, 100876, China}

\begin{abstract}
  We introduce a hierarchy of tests satisfied by any probability distribution $P$ that represents the quantum correlations generated in prepare-and-measure (P\&M) quantum multi-chain-shaped networks, assuming only the inner-product information of prepared states. The P\&M quantum multi-chain-shaped networks involve multiple measurement parties, with each measurement party potentially having multiple sequential receivers. We adapt the original NPA-hierarchy by incorporating a finite number of linear and positive semi-definite constraints to characterize the quantum correlations in P\&M quantum multi-chain-shaped networks. These constraints in each hierarchy are derived from sequential measurements and the inner-product matrix of prepared states. The adapted NPA-hierarchy is further applied to tackle some quantum information tasks, including sequential quantum random access codes (QRACs) and semi-device-independent randomness certification. First, we derive the optimal trade-off between the two sequential receivers in the $2 \to 1$ sequential QRACs, and investigate randomness certification in its double violation region. Second, considering the presence of an eavesdropper in actual communication, we show how much local and global randomness can be certified using the optimal trade-off of $2 \to 1$ sequential QRACs. We also quantify the amount of local and global randomness that can be certified from the complete set of probabilities generated by the two sequential receivers. Our conclusion is that utilizing the full set of probabilities certifies more randomness than relying solely on the optimal trade-off relationship.  
\end{abstract}
\maketitle

 \section{Introduction}
The correlations generated in quantum systems exhibit nonclassical behavior, offering unique advantages over data obtained from classical sources. Over the past decades, physicists have devoted their efforts to studying quantum correlations generated by various quantum resources, such as entanglement 
 \cite{RevModPhys.81.865}, steering \cite{PhysRevLett.98.140402}, nonlocality \cite{RevModPhys.86.419,Cavalcanti_2011}, and contextuality \cite{PhysRevA.71.052108}. These distinct quantum correlations serve as the essential building blocks of quantum theory. Additionally, optimizing over the entire set of quantum correlations has revealed some intriguing applications, such as randomness certification \cite{PhysRevLett.108.100402, PhysRevA.88.052116} and quantum random access codes (QRACs) \cite{e92-a_5_1268}. Quantum correlations are fundamental to quantum information science, and their characterization represents a significant challenge in quantum information theory.
 
The direct strategy to characterize quantum correlations is searching over all quantum states and measurements, which is clearly infeasible. Currently, the most effective method for addressing this problem is the NPA-hierarchy, which characterizes the set of quantum correlations in the standard Bell scenario through a sequence of increasingly tighter outer approximations, each formulated as a semi-definite program \cite{PhysRevLett.98.010401,Navascués_2008}. With the advancement of quantum technology, causal quantum networks \cite{Fritz_2012} going beyond the standard Bell scenario have been developed. Subsequently, more complex quantum Bell networks have attracted much attention, such as star-shaped quantum networks \cite{9571334,Andreoli_2017,PhysRevA.104.042217,PhysRevA.90.062109}, tree-shaped quantum networks \cite{Yang:2023ylc,PhysRevA.110.012401} and multi-chain-shaped quantum networks \cite{Gallego_2014,PhysRevA.108.042409}. The research in Ref. \cite{PhysRevLett.123.140503} adapted the NPA-hierarchy to bound the set of quantum correlations in causal quantum networks by imposing relaxations of factorization constraints in a form compatible with semi-definite programming. Besides the Bell scenario, prepare-and-measure (P\&M) is another common scenario in quantum information processing. Bowles et al. firstly used dimension witnesses to certify nonclassicality correlations in P\&M scenario \cite{PhysRevA.92.022351}. Later, the work in Ref. \cite{Wang2019} adapted the original NPA-hierarchy by adding certain linear constraints to characterize the correlations in P\&M quantum networks involving multiple measurement parties. 
\begin{figure*}[ht!] 
	\centering
	\begin{subfigure}[b]{0.48\textwidth} 
		\centering
		\includegraphics[width=\textwidth]{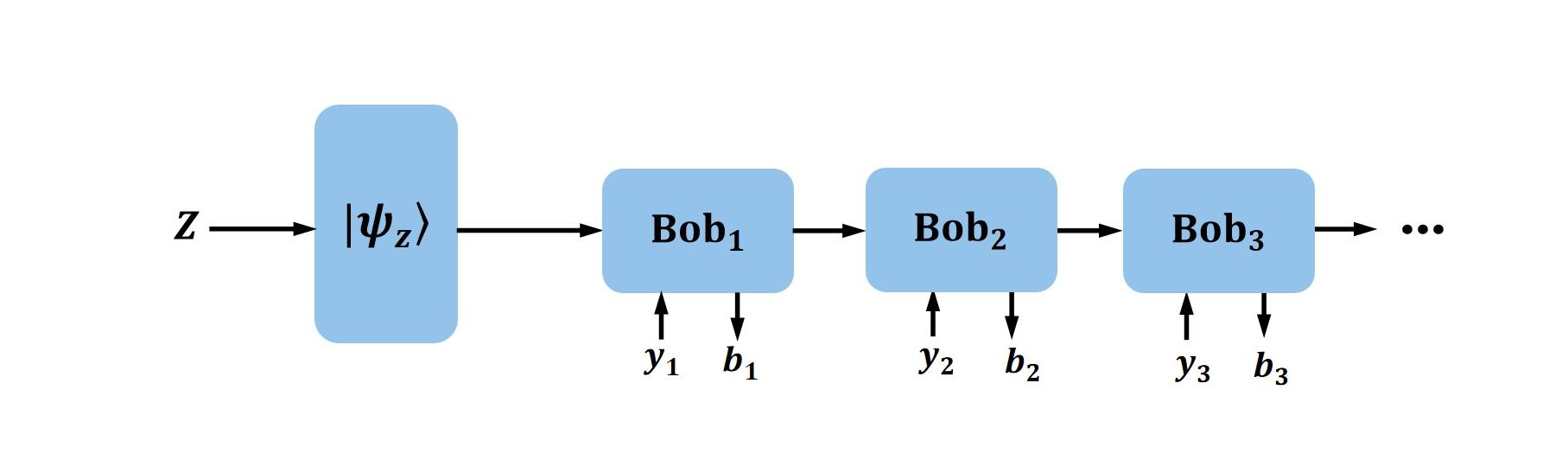} 
		\caption{\centering}
		\label{fig:subfigure1}
	\end{subfigure}
	\hfill
	\begin{subfigure}[b]{0.48\textwidth} 
		\centering
		\includegraphics[width=\textwidth]{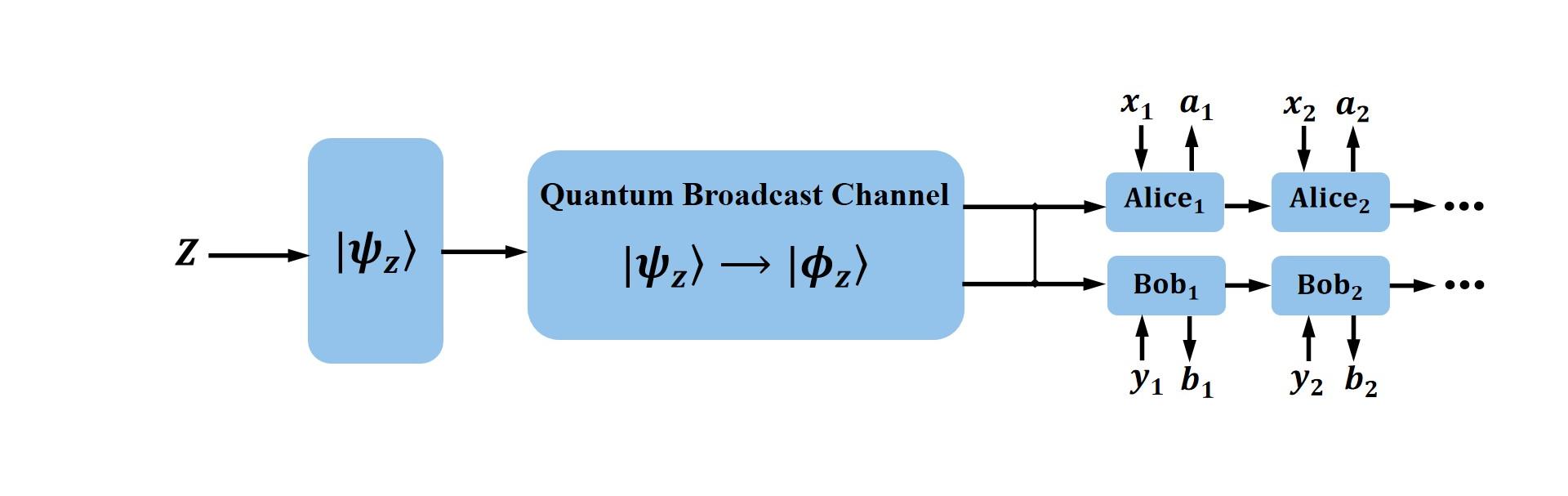} 
		\caption{\centering}
		\label{fig:subfigure2}
	\end{subfigure}
    \caption{(a) The P\&M quantum one-chain-shaped network with a single measurement party involving multiple sequential receivers. (b) The P\&M quantum two-chain-shaped network with two measurement parties, each involving multiple sequential receivers.}
	\label{Fig:1} 
\end{figure*} 

Sequential measurements \cite{Silva_2015} offer advantages in certain quantum information tasks. For example, employing sequences of measurements can overcome the limitations on the amount of randomness in the standard Bell scenario \cite{PhysRevA.95.020102,Liu_2024}. Furthermore, several studies have demonstrated that sequential measurements enable more receivers to simultaneously exhibit distinct quantum correlations \cite{PhysRevLett.125.090401,Hu2018,PhysRevA.98.012305,PhysRevA.98.062304,PhysRevA.98.042311}. These findings suggest potential applications for sequential measurements in quantum networks. Bowles et al. effectively adapted the original NPA-hierarchy by incorporating finite linear constraints to characterize the quantum correlations generated in the sequential Bell scenario \cite{Bowles2020boundingsetsof}. Sequential measurements also can be applied in P\&M quantum scenarios \cite{Mohan_2019}. In practical communication, quantum devices encounter challenges posed by a potential eavesdropper Eve \cite{Gerhardt2011,PhysRevApplied.15.034081,Wath2023}. In sequential P\&M quantum scenario, the presence of Eve implies that the prepared states are transmitted to two measurement parties. We extend our consideration to more general scenarios, namely P\&M quantum multi-chain-shaped networks. These networks involve an arbitrary number of measurement parties, each of which may have multiple sequential receivers. 

In this paper, we concrete on characterizing the quantum correlations in P\&M quantum multi-chain-shaped networks. The sequential measurements in this scenario are conducted on a varying prepared state $|\psi_z\rangle$ determined by the classical input $z$ rather than a fixed state $|\psi\rangle$. The states in $\{|\psi_z\rangle\}_{z=1}^n$ are non-orthogonal. These varying and non-orthogonal quantum states are indistinguishable. The inner-product information of prepared states plays a crucial role in verifying the security of quantum cryptographic protocol \cite{PhysRevA.99.062332,Pereira2019,PhysRevApplied.12.024048}. To characterize the quantum correlations in P\&M quantum multi-chain-shaped networks, we adapt the original NPA-hierarchy by augmenting it with a finite number of linear constraints. Specifically, we introduce a sequence of necessary conditions for the quantum correlations set, assuming only the inner-product information of the prepared states. Each of our conditions amounts to verifying the existence of a positive semi-definite matrix that must satisfy certain linear constraints. These constraints stem from the inner-products of prepared states and the properties of sequential measurements. If one of our conditions is not satisfied, we can immediately conclude that the given correlation is not quantum.  

To demonstrate the feasibility and efficacy of our method, we apply our adapted hierarchy to several quantum information tasks. First, we derive the optimal trade-off between the two sequential receivers in $2 \to 1$ sequential QRACs, replicating the results from Ref. \cite{Mohan_2019}. We also investigate randomness certification based on the double violation region in $2 \to 1$ sequential QRACs. Our approach relaxes the dimension assumptions compared to the general semi-device-independent (SDI) randomness certification using $2 \to 1$ QRACs \cite{PhysRevA.84.034301,PhysRevA.85.052308,PhysRevA.92.022331}, requiring only the prepared states' inner-product information. Second, in the presence of an eavesdropper Eve, we analyze how much local and global randomness can be certified using the optimal trade-off of $2 \to 1$ sequential QRACs. We also derive the certified local and global randomness even when Eve knows the complete observed probability distribution $P_{obs}$ generated by two sequential receivers. We establish the relationship between the amount of certified randomness and the measurement sharpness parameter of the first sequential receiver. The results show that the full set of probabilities certifies more randomness than relying solely on the optimal trade-off relationship.
\section{The P\&M quantum multi-chain-shaped networks}
The simplest P\&M quantum multi-chain-shaped network is the standard sequential P\&M quantum scenario, which involves one measurement party with sequential receivers as illustrated in Fig.~\ref{Fig:1}(a). A classical random source $z$ is encoded into a quantum system $|\psi_z\rangle$ and distributed to Bob$_1$ for measurement via a quantum channel. The encoding states in $\{|\psi_z\rangle\}_{z=1}^n$ are non-orthogonal. Then, Bob$_1$ passes the post-measurement state to subsequent sequential receivers who repeat this process. In the general P\&M quantum multi-chain-shaped network with more than one measurement party, the prepared states will be transmitted to them through a quantum broadcast channel. We will focus our discussion on a P\&M quantum multi-chain-shaped network limited to two measurement parties with sequential receivers, namely P\&M quantum two-chain-shaped network, as illustrated in Fig. \ref{Fig:1}(b). Our analysis can be straightforwardly extended to the general networks involving an arbitrary number of measurement parties with sequential receivers.

A classical random source $z$ $(z \in \{1, 2, ..., n\})$ is encoded into a quantum state $|\psi_z\rangle$. The encoding states in $\{|\psi_z\rangle\}_{z=1}^n$ are non-orthogonal. Subsequently, one of the $n$ predefined states ${|\psi_z\rangle}_{z=1}^n$, denoted as $|\psi_z\rangle$, is converted into $|\phi_z\rangle$ after being transmitted through the quantum broadcast channel. Then, $|\phi_z\rangle$ is distributed to $\text{Alice}_1$ and $\text{Bob}_1$. For this type of transmission, adopting the purification picture as described in Ref. \cite{Wilde_2017}, we may see the state transformation is described by unitary evolution. Specifically, by working in a higher-dimensional Hilbert space, the isometric evolution takes $|\psi_z\rangle$ to some pure state $|\phi_z\rangle$. After the transmission, the dimension and other properties of $|\psi_z\rangle$ may change, while the initial information about the inner-product is preserved in the transformed states, i.e., $\langle\phi_z|\phi_{z^\prime} \rangle = \langle\psi_z|\psi_{z^\prime} \rangle= \lambda_{zz^\prime}$. After $\text{Alice}_1$ and $\text{Bob}_1$ have performed their randomly selected measurements $x_1,y_1$ and recorded the outcomes $a_1,b_1$, they respectively pass the post-measurement particles to $\text{Alice}_2$ and $\text{Bob}_2$ who repeat this process.

Without loss of generality, we assume that there are $m$ sequential receivers at Alice's side and Bob's side, respectively. The
entire sequence at Alice's side has inputs $\mathbf{x} = (x_1,x_2,\cdots,x_m)$ and outputs $\mathbf{a} = (a_1,a_2,\cdots,a_m)$. We denote the sequential measurements at Alice's side as in Ref. \cite{Bowles2020boundingsetsof}:
   \begin{align}
	&&A_{\mathbf{x}}^{\mathbf{a}}=\int{.}..\int{d}\mu _1...d\mu _m{K_{x_1}^{a_1,\mu _1}}^\dagger {K_{x_2}^{a_2,\mu _2}}^\dagger \notag \\
	&&...{K_{x_m}^{a_m,\mu _m}}^\dagger K_{x_m}^{a_m,\mu _m}...K_{x_2}^{a_2,\mu _2}K_{x_1}^{a_1,\mu _1},
	\label{eq:1}
\end{align}
where the $i$th measurement input and its corresponding outcome characterized by sets of Kraus operators $\{K_{x_i}^{a_i,\mu_i}\}$. Here, the index $\mu_i$ indicates that there may be multiple Kraus operators associated with a single measurement outcome $a_i$. The Kraus operator satisfy $\sum_{a_i}\int d\mu_i{K_{x_i}^{a_i,\mu_i}}^\dagger K_{x_i}^{a_i,\mu_i} = I_A$. We denote the sequential inputs and outputs at Bob's side by $\mathbf{y} = (y_1,y_2,\cdots,y_m)$ and $\mathbf{b} = (b_1,b_2,\cdots,b_m)$, respectively. The sequential measurement $B_\mathbf{y}^\mathbf{b}$ has a similar structure to $A_\mathbf{x}^\mathbf{a}$. 

 Using the quantum Born rule, we have that the probability of observing outcomes $\mathbf{a},\mathbf{b}$ given sequential measurements $\mathbf{x},\mathbf{y}$ and the set $\{|\phi_z\rangle\}_{z=1}^n$ is
\begin{align}
	p(\mathbf{a},\mathbf{b}|\mathbf{x},\mathbf{y},z)= \langle\phi_z|A_\mathbf{x}^\mathbf{a}B_\mathbf{y}^\mathbf{b}|\phi_z \rangle,
	\label{eq:2}
\end{align}
where $A_\mathbf{x}^\mathbf{a}$ and $B_\mathbf{y}^\mathbf{b}$ follow the sequential structure in Eq. \eqref{eq:1}. We denote a set of quantum correlations that generated by the P\&M multi-chain-shaped networks as $Q(\lambda)^{SEQ}$, where $\lambda$ is the inner-product matrix of prepared states. A given set of correlations $p(\mathbf{a},\mathbf{b}|\mathbf{x},\mathbf{y},z)$ belongs to $Q(\lambda)^{SEQ}$ if and only if it can be realised as in Eq. \eqref{eq:2}, with the measurements satisfying the following properties \cite{Bowles2020boundingsetsof}:
    \begin{align}
(i)\quad& A_\mathbf{x}^\mathbf{a}A_\mathbf{x}^{\mathbf{a}^{\prime}} = \delta_{\mathbf{a},\mathbf{a}^{\prime}}A_\mathbf{x}^\mathbf{a}, \quad \forall \mathbf{x},\mathbf{a},\mathbf{a}^{\prime}, \notag \\
(ii)\quad&\sum_{a_{k+1},\cdots,a_m}A_\mathbf{x}^\mathbf{a}-A_{\mathbf{x}^{\prime}}^\mathbf{a} = 0, \quad \forall a_1,a_2,\cdots,a_k, \forall \mathbf{x},\mathbf{x}^{\prime},\notag\\
& s.t.\quad x_i = x_i^{\prime},(i\leq k),\notag \\ 
& \qquad1\leq k\leq m-1,\notag \\
(iii)\quad& A_{\mathbf{x}}^{\mathbf{a}}A_{\mathbf{x}^{\prime}}^{\mathbf{a}^{\prime}}=0,\quad \forall \mathbf{x}, \mathbf{x}^{\prime}, \mathbf{a}, \mathbf{a}^{\prime},\notag \\
&s.t.\quad x_i = x_i^{\prime},(i\leq k), \notag \\
& \qquad(a_1,\cdots,a_k)\neq(a_1^{\prime},\cdots,a_k^{\prime}),\notag \\
& \qquad 1\leq k\leq m,\notag \\
(iv)\quad&[A_\mathbf{x}^\mathbf{a},B_\mathbf{y}^\mathbf{b}]=0,
\label{eq:3}
\end{align}
and similarly $(i)$-$(iii)$ for $B_\mathbf{y}^\mathbf{b}$. Properties $(i)$ and $(iii)$ can be deduced from the construction of the sequential measurements. Property $(ii)$ represents the one-way ‘no-signaling’ conditions because the sequential measurements that define the first $k$ measurements must
be independent of the last $m-k$ inputs. Property $(iv)$ indicates that the sequential measurements on Alice's side and Bob's side commute with each other.
\section{The adapted NPA-hierachy for P\&M two-chain-shaped networks}
The problem that we aim to tackle is the following: Given an arbitrary probability distribution $P$ and an $n \times n$  matrix $\lambda$, do there exist some quantum states $\{|\phi_z\rangle\}_{z=1}^{n}$ and measurements $A_\mathbf{x}^\mathbf{a}$, $B_\mathbf{y}^\mathbf{b}$, such that $p(\mathbf{a},\mathbf{b}|\mathbf{x},\mathbf{y},z)= \langle\phi_z|A_\mathbf{x}^\mathbf{a}B_\mathbf{y}^\mathbf{b}|\phi_z \rangle$? The inner-product of prepared states satisfies $\langle\phi_z|\phi_{z^{\prime}}\rangle = \lambda_{zz^{\prime}}$, $z,z^{\prime} \in \{1,2,\cdots,n\}$. The measurements $A_\mathbf{x}^\mathbf{a}$ and $B_\mathbf{y}^\mathbf{b}$ satisfy the properties in Eq. \eqref{eq:3}.

The adapted NPA-hierarchy can efficiently characterize the set of quantum correlations for any P\&M quantum multi-chain-shaped network, assuming only the inner-product matrix $\lambda$. Let $S_k = \{S_k^1, S_k^2,...,S_k^l\}$ be a set of $l$ operators in level $k$, where each element $S_k^i$ is the identity operator or a linear combination of products of $A_\mathbf{x}^\mathbf{a}$ and $B_\mathbf{y}^\mathbf{b}$. More precisely, we define a sequence of hierarchical sets:
\begin{align}
	&S_1 = \{I\} \cup_{\mathbf{a},\mathbf{x}}\{A_\mathbf{x}^\mathbf{a}\} \cup_{\mathbf{b},\mathbf{y}}\{B_\mathbf{y}^\mathbf{b}\},  \notag
    \\& S_{k+1} = S_k \cup_{i,j}\{S_k^iS_1^j\}. 
    \label{eq:4}
\end{align}

Then define $G$ to be an $nl \times nl$ block matrix
\begin{align}
G=\sum_{z,z^{\prime} =1}^{n}G^{zz^{\prime}}\otimes|e_z\rangle\langle e_{z^{\prime}}|,
\label{eq:5}
\end{align}
 where $G^{zz^{\prime}}$ is an $l \times l$ matrix with the $ij-$entry as $G^{zz^{\prime}}_{\left( i,j \right)} = \langle \phi_z|{{S_k^i}}^\dagger S_k^j|\phi_{z^{\prime}}\rangle, \forall i,j\in \{1,2,...,l\}, \forall z,z^{\prime} \in \{1,2,...,n\}$, and $\{|e_z\rangle\}_{z=1}^n$ is the standard orthonormal basis of $\mathbb{C}^n$.  By construction, the matrix $G$ is positive semi-definite. Furthermore, the properties of sequential measurements and the inner-product constraints $\langle\phi_z|\phi_{z^{\prime}}\rangle = \lambda_{zz^{\prime}}$ translate to linear constraints on the entries of $G$:
\begin{itemize}
	\item[$(i)$] 
	$G$ satisfies a number of linear constraints that arise from sequential measurements properties, as described in Eq. \eqref{eq:3}. These constraints can be represented as $tr[G^{zz^\prime}M_k]=0$ and $tr[G^{zz^\prime}M_{k}^{SEQ}]=0$ with appropriately chosen fixed matrices $M_k$ and $M_{k}^{SEQ}$.
	\item [$(ii)$]
	$G$ includes elements that directly correspond to the joint probabilities $p(\mathbf{a},\mathbf{b}|\mathbf{x},\mathbf{y},z)= \langle\phi_z|A_\mathbf{x}^\mathbf{a}B_\mathbf{y}^\mathbf{b}|\phi_z \rangle$. We write these constraints as $tr[G^{zz}F_k]=P^{zz}$ , where $F_k$ is fixed matrix and $P^{zz}$ denotes the corresponding joint probabilities.
	\item [$(iii)$] 
	By setting $S_k^{(1)}=I, \forall k$, we have $G_{\left( 1,1 \right)}^{zz^\prime}=\lambda _{zz^\prime}$.
\end{itemize}

If the given correlations $P\in Q(\lambda)^{SEQ}$, there exist some quantum states and sequential measurements leading to $P$, along with a corresponding matrix $G$ that satisfies the above constraints. We outline a crucial necessary condition for $P\in Q(\lambda)^{SEQ}$ as follows. If we define the operator sets as $S_k$, then the set of correlations that satisfy the positive solution in the adapted NPA-hierarchy (level $k$) is denoted as $Q_k(\lambda)^{SEQ}$.

The adapted NPA-hierarchy (level $k$):
\begin{align}
    &\text{Find} \quad G
    \notag
	\\ \text{subject to:}\quad &G\ge 0, G^\dagger=G, G_{\left( 1,1 \right)}^{zz^{\prime}}=\lambda _{zz^{\prime}},
	\notag
	\\&tr[G^{zz^\prime}M_k]=0,
	\notag
	\\&tr[G^{zz^\prime}M_{k}^{SEQ}]=0,
	\notag
	\\&tr[G^{zz}F_k]=P^{zz}.
	\label{eq:6}
\end{align}

The problem is characterized by linear and positive semi-definite constraints, making it suitable for formulation as a SDP feasibility problem. The computational complexity of the solution depends on the size of the matrix $G$. Since this test represents a necessary condition for membership in $ Q(\lambda)^{SEQ} $, it implies that $Q(\lambda)^{SEQ}$ is a subset of $Q_k(\lambda)^{SEQ}$. Consequently, we derive a sequence of SDPs, each providing a relaxation towards determining membership in $Q(\lambda)^{SEQ}$.

Given that the properties in Eq. \eqref{eq:3} characterize precisely the set of sequential measurement operators and  establish linear constraints, methods similar to those in Ref. \cite{Wang2019} can be employed to demonstrate convergence of the hierarchy. As the set $\{S_k\}$ satisfying $S_k \subseteq S_{k+1}$, the sequence of block matrix, $G_1, G_2, \ldots$, has increasing size and constraints. The growing hierarchy offers a progressively tighter approximation of the quantum set: $Q(\lambda)^{SEQ} \subseteq Q_k(\lambda)^{SEQ} \subseteq Q_{k-1}(\lambda)^{SEQ} \cdots$. It converges to the quantum correlation sets, i.e $\lim\limits_{k \to \infty}Q_k(\lambda)^{SEQ} = Q(\lambda)^{SEQ}$. One can extract some quantum states and sequential measurements from the matrix $G_\infty$ corresponding to the asymptotic level of the hierarchy.
\section{The Applications}
The method described above can effectively tackle some information processing tasks in P\&M quantum multi-chain-shaped networks, including sequential QRACs and SDI randomness certification. Here we provide four concrete applications in two simple P\&M quantum multi-chain-shaped networks.
\subsection{The P\&M quantum one-chain-shaped network}
\subsubsection{The optimal trade-off of $2 \rightarrow 1$ sequential QRACs}
\begin{figure}
	\includegraphics[width=0.5\textwidth]{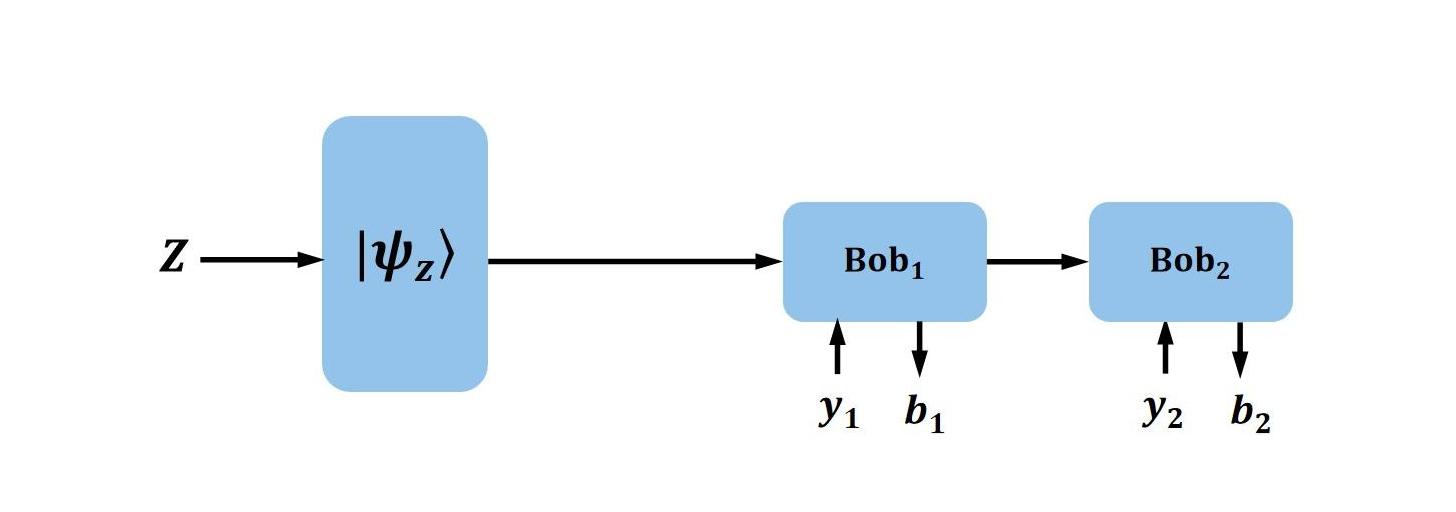}
	\caption{The P\&M quantum one-chain-shaped network with two sequential receivers. } 
	\label{Fig2} 
\end{figure}

QRACs are key tools in quantum information theory, particularly in studying information capacity and the encoding efficiency of quantum states. They are used to encode classical bit strings into quantum states, enabling the recovery of the original bit strings with a certain probability in subsequent processes \cite{e92-a_5_1268,PhysRevLett.114.170502}. The Ref. \cite{Mohan_2019} introduced sequential QRACs beyond standard QRACs, thereby enhancing access to quantum system information for sequential receivers. The P\&M quantum one-chain-shaped network involves two sequential receivers can be used to implement $2 \rightarrow 1$ sequential QRACs (see Fig. \ref{Fig2}).

In a $2 \rightarrow 1$ sequential QRACs, the two random classical bits $z_0z_1$ ($\{z_0,z_1\} \in \{0,1\}^2$)  are encoded into a quantum state $|\phi_z\rangle$. We assume nothing about these states, except that the inner-product matrix $\lambda$ of the code states $\{|\phi_{00}\rangle,|\phi_{01}\rangle,|\phi_{10}\rangle,|\phi_{11}\rangle\}$ is fixed to that of $\left\lbrace |0\rangle,|+\rangle,|-\rangle,|1\rangle \right\rbrace$. Thus,
\begin{align}
	\lambda = \begin{bmatrix}
		1 & \frac{1}{\sqrt{2}} & \frac{1}{\sqrt{2}} & 0 \\
		\frac{1}{\sqrt{2}} & 1 & 0 & -\frac{1}{\sqrt{2}} \\
		\frac{1}{\sqrt{2}} & 0 & 1 & \frac{1}{\sqrt{2}} \\
		0 & -\frac{1}{\sqrt{2}} & \frac{1}{\sqrt{2}} & 1 
	\end{bmatrix}.
	\label{eq:7}
\end{align}
\begin{figure*}[ht!]
	\centering
	\[
	\renewcommand{\arraystretch}{3}
	\scalebox{0.46}{$
		\left(
		\begin{array}{c|c|c|c|c|c|c|c|c|c|c|c|c|c|c|c|c}
			\lambda_{zz^{\prime}} & \langle \phi_z |B_{00}^{00}| \phi_{z^{\prime}} \rangle & \langle \phi_z |B_{00}^{01}| \phi_{z^{\prime}} \rangle & \langle \phi_z |B_{00}^{10}| \phi_{z^{\prime}} \rangle & \langle \phi_z |B_{00}^{11}| \phi_{z^{\prime}} \rangle & \langle \phi_z |B_{01}^{00}| \phi_{z^{\prime}} \rangle & \langle \phi_z |B_{01}^{01}| \phi_{z^{\prime}} \rangle & \langle \phi_z |B_{01}^{10}| \phi_{z^{\prime}} \rangle & \langle \phi_z |B_{01}^{11}| \phi_{z^{\prime}} \rangle & \langle \phi_z |B_{10}^{00}| \phi_{z^{\prime}} \rangle & \langle \phi_z |B_{10}^{01}| \phi_{z^{\prime}} \rangle & \langle \phi_z |B_{10}^{10}| \phi_{z^{\prime}} \rangle & \langle \phi_z |B_{10}^{11}| \phi_{z^{\prime}} \rangle & \langle \phi_z |B_{11}^{00}| \phi_{z^{\prime}} \rangle & \langle \phi_z |B_{11}^{01}| \phi_{z^{\prime}} \rangle & \langle \phi_z |B_{11}^{10}| \phi_{z^{\prime}} \rangle & \langle \phi_z |B_{11}^{11}| \phi_{z^{\prime}} \rangle \\
			\hline
			* & \langle \phi_z |B_{00}^{00}| \phi_{z^{\prime}} \rangle & 0 & 0 & 0 & \langle \phi_z |B_{00}^{00}B_{01}^{00}| \phi_{z^{\prime}} \rangle & \langle \phi_z |B_{00}^{00}B_{01}^{01}| \phi_{z^{\prime}} \rangle & 0 & 0 & \langle \phi_z |B_{00}^{00}B_{10}^{00}| \phi_{z^{\prime}} \rangle & \langle \phi_z |B_{00}^{00}B_{10}^{01}| \phi_{z^{\prime}} \rangle & \langle \phi_z |B_{00}^{00}B_{10}^{10}| \phi_{z^{\prime}} \rangle & \langle \phi_z |B_{00}^{00}B_{10}^{11}| \phi_{z^{\prime}} \rangle & \langle \phi_z |B_{00}^{00}B_{11}^{00}| \phi_{z^{\prime}} \rangle & \langle \phi_z |B_{00}^{00}B_{11}^{01}| \phi_{z^{\prime}} \rangle & \langle \phi_z |B_{00}^{00}B_{11}^{10}| \phi_{z^{\prime}} \rangle & \langle \phi_z |B_{00}^{00}B_{11}^{11}| \phi_{z^{\prime}} \rangle \\
			\hline
			* & 0 & \langle \phi_z |B_{00}^{01}| \phi_{z^{\prime}} \rangle & 0 & 0 & \langle \phi_z |B_{00}^{01}B_{01}^{00}| \phi_{z^{\prime}} \rangle & \langle \phi_z |B_{00}^{01}B_{01}^{01}| \phi_{z^{\prime}} \rangle & 0 & 0 & \langle \phi_z |B_{00}^{01}B_{10}^{00}| \phi_{z^{\prime}} \rangle & \langle \phi_z |B_{00}^{01}B_{10}^{01}| \phi_{z^{\prime}} \rangle & \langle \phi_z |B_{00}^{01}B_{10}^{10}| \phi_{z^{\prime}} \rangle & \langle \phi_z |B_{00}^{01}B_{10}^{11}| \phi_{z^{\prime}} \rangle & \langle \phi_z |B_{00}^{01}B_{11}^{00}| \phi_{z^{\prime}} \rangle & \langle \phi_z |B_{00}^{01}B_{11}^{01}| \phi_{z^{\prime}} \rangle & \langle \phi_z |B_{00}^{01}B_{11}^{10}| \phi_{z^{\prime}} \rangle & \langle \phi_z |B_{00}^{01}B_{11}^{11}| \phi_{z^{\prime}} \rangle  \\
			\hline
			* & 0 & 0 & \langle \phi_z |B_{00}^{10}| \phi_{z^{\prime}} \rangle & 0 & 0 & 0 & \langle \phi_z |B_{00}^{10}B_{01}^{10}| \phi_{z^{\prime}} \rangle & \langle \phi_z |B_{00}^{10}B_{01}^{11}| \phi_{z^{\prime}} \rangle & \langle \phi_z |B_{00}^{10}B_{10}^{00}| \phi_{z^{\prime}} \rangle & \langle \phi_z |B_{00}^{10}B_{10}^{01}| \phi_{z^{\prime}} \rangle & \langle \phi_z |B_{00}^{10}B_{10}^{10}| \phi_{z^{\prime}} \rangle & \langle \phi_z |B_{00}^{01}B_{10}^{11}| \phi_{z^{\prime}} \rangle & \langle \phi_z |B_{00}^{10}B_{11}^{00}| \phi_{z^{\prime}} \rangle & \langle \phi_z |B_{00}^{10}B_{11}^{01}| \phi_{z^{\prime}} \rangle & \langle \phi_z |B_{00}^{10}B_{11}^{10}| \phi_{z^{\prime}} \rangle & \langle \phi_z |B_{00}^{10}B_{11}^{11}| \phi_{z^{\prime}} \rangle \\
			\hline
			* & 0 & 0 & 0 & \langle \phi_z |B_{00}^{11}| \phi_{z^{\prime}} \rangle & 0 & 0 & \langle \phi_z |B_{00}^{11}B_{01}^{10}| \phi_{z^{\prime}} \rangle & \langle \phi_z |B_{00}^{11}B_{01}^{11}| \phi_{z^{\prime}} \rangle & \langle \phi_z |B_{00}^{11}B_{10}^{00}| \phi_{z^{\prime}} \rangle & \langle \phi_z |B_{00}^{11}B_{10}^{01}| \phi_{z^{\prime}} \rangle & \langle \phi_z |B_{00}^{11}B_{10}^{10}| \phi_{z^{\prime}} \rangle & \langle \phi_z |B_{00}^{11}B_{10}^{11}| \phi_{z^{\prime}} \rangle & \langle \phi_z |B_{00}^{11}B_{11}^{00}| \phi_{z^{\prime}} \rangle & \langle \phi_z |B_{00}^{11}B_{11}^{01}| \phi_{z^{\prime}} \rangle & \langle \phi_z |B_{00}^{11}B_{11}^{10}| \phi_{z^{\prime}} \rangle & \langle \phi_z |B_{00}^{11}B_{11}^{11}| \phi_{z^{\prime}} \rangle \\
			\hline
			* & * & * & 0 & 0 & \langle \phi_z |B_{01}^{00}| \phi_{z^{\prime}} \rangle & 0 & 0 & 0 & \langle \phi_z |B_{01}^{00}B_{10}^{00}| \phi_{z^{\prime}} \rangle & \langle \phi_z |B_{01}^{00}B_{10}^{01}| \phi_{z^{\prime}} \rangle & \langle \phi_z |B_{01}^{00}B_{10}^{10}| \phi_{z^{\prime}} \rangle & \langle \phi_z |B_{01}^{00}B_{10}^{11}| \phi_{z^{\prime}} \rangle & \langle \phi_z |B_{01}^{00}B_{11}^{00}| \phi_{z^{\prime}} \rangle & \langle \phi_z |B_{01}^{00}B_{11}^{01}| \phi_{z^{\prime}} \rangle & \langle \phi_z |B_{01}^{00}B_{11}^{10}| \phi_{z^{\prime}} \rangle & \langle \phi_z |B_{01}^{00}B_{11}^{11}| \phi_{z^{\prime}} \rangle \\
			\hline
			* & * & * & 0 & 0 & 0 & \langle \phi_z |B_{01}^{01}| \phi_{z^{\prime}} \rangle & 0 & 0 & \langle \phi_z |B_{01}^{01}B_{10}^{00}| \phi_{z^{\prime}} \rangle & \langle \phi_z |B_{01}^{01}B_{10}^{01}| \phi_{z^{\prime}} \rangle & \langle \phi_z |B_{01}^{01}B_{10}^{10}| \phi_{z^{\prime}} \rangle & \langle \phi_z |B_{01}^{01}B_{10}^{11}| \phi_{z^{\prime}} \rangle & \langle \phi_z |B_{01}^{01}B_{11}^{00}| \phi_{z^{\prime}} \rangle & \langle \phi_z |B_{01}^{01}B_{11}^{01}| \phi_{z^{\prime}} \rangle & \langle \phi_z |B_{01}^{01}B_{11}^{10}| \phi_{z^{\prime}} \rangle & \langle \phi_z |B_{01}^{01}B_{11}^{11}| \phi_{z^{\prime}} \rangle \\
			\hline
			* & 0 & 0 & * & * & 0 & 0 & \langle \phi_z |B_{01}^{10}| \phi_{z^{\prime}} \rangle & 0 & \langle \phi_z |B_{01}^{10}B_{10}^{00}| \phi_{z^{\prime}} \rangle & \langle \phi_z |B_{01}^{10}B_{10}^{01}| \phi_{z^{\prime}} \rangle & \langle \phi_z |B_{01}^{10}B_{10}^{10}| \phi_{z^{\prime}} \rangle & \langle \phi_z |B_{01}^{10}B_{10}^{11}| \phi_{z^{\prime}} \rangle & \langle \phi_z |B_{01}^{10}B_{11}^{00}| \phi_{z^{\prime}} \rangle & \langle \phi_z |B_{01}^{10}B_{11}^{01}| \phi_{z^{\prime}} \rangle & \langle \phi_z |B_{01}^{10}B_{11}^{10}| \phi_{z^{\prime}} \rangle & \langle \phi_z |B_{01}^{10}B_{11}^{11}| \phi_{z^{\prime}} \rangle \\
			\hline
			* & 0 & 0 & * & * & 0 & 0 & 0 & \langle \phi_z |B_{01}^{11}| \phi_{z^{\prime}} \rangle & \langle \phi_z |B_{01}^{11}B_{10}^{00}| \phi_{z^{\prime}} \rangle & \langle \phi_z |B_{01}^{11}B_{10}^{01}| \phi_{z^{\prime}} \rangle & \langle \phi_z |B_{01}^{11}B_{10}^{10}| \phi_{z^{\prime}} \rangle & \langle \phi_z |B_{01}^{11}B_{10}^{11}| \phi_{z^{\prime}} \rangle & \langle \phi_z |B_{01}^{11}B_{11}^{00}| \phi_{z^{\prime}} \rangle & \langle \phi_z |B_{01}^{11}B_{11}^{01}| \phi_{z^{\prime}} \rangle & \langle \phi_z |B_{01}^{11}B_{11}^{10}| \phi_{z^{\prime}} \rangle & \langle \phi_z |B_{01}^{11}B_{11}^{11}| \phi_{z^{\prime}} \rangle \\
			\hline
			* & * & * & * & * & * & * & * & * & \langle \phi_z |B_{10}^{00}| \phi_{z^{\prime}} \rangle & 0 & 0 & 0 & \langle \phi_z |B_{10}^{00}B_{11}^{00}| \phi_{z^{\prime}} \rangle & \langle \phi_z |B_{10}^{00}B_{11}^{01}| \phi_{z^{\prime}} \rangle & 0 & 0 \\
			\hline
			* & * & * & * & * & * & * & * & * & 0 & \langle \phi_z |B_{10}^{01}| \phi_{z^{\prime}} \rangle & 0 & 0 & \langle \phi_z |B_{10}^{01}B_{11}^{00}| \phi_{z^{\prime}} \rangle & \langle \phi_z |B_{10}^{01}B_{11}^{01}| \phi_{z^{\prime}} \rangle & 0 & 0 \\
			\hline
			* & * & * & * & * & * & * & * & * & 0 & 0 & \langle \phi_z |B_{10}^{10}| \phi_{z^{\prime}} \rangle & 0 & 0 & 0 & \langle \phi_z |B_{10}^{10}B_{11}^{10}| \phi_{z^{\prime}} \rangle & \langle \phi_z |B_{10}^{10}B_{11}^{11}|\phi_{z^{\prime}} \rangle \\
			\hline
			* & * & * & * & * & * & * & * & * & 0 & 0 & 0 & \langle \phi_z |B_{10}^{11}| \phi_{z^{\prime}} \rangle & 0 & 0 & \langle \phi_z |B_{10}^{11}B_{11}^{10}| \phi_{z^{\prime}} \rangle & \langle \phi_z |B_{10}^{11}B_{11}^{11}|\phi_{z^{\prime}} \rangle \\
			\hline
			* & * & * & * & * & * & * & * & * & * & * & 0 & 0 & \langle \phi_z |B_{11}^{00}| \phi_{z^{\prime}} \rangle & 0 & 0 & 0 \\
			\hline
			* & * & * & * & * & * & * & * & * & * & * & 0 & 0 & 0 & \langle \phi_z |B_{11}^{01}| \phi_{z^{\prime}} \rangle & 0 & 0 \\
			\hline
			* & * & * & * & * & * & * & * & * & 0 & 0 & * & * & 0 & 0 & \langle \phi_z |B_{11}^{10}| \phi_{z^{\prime}} \rangle & 0 \\
			\hline
			* & * & * & * & * & * & * & * & * & 0 & 0 & * & *& 0 & 0 & 0 & \langle \phi_z |B_{11}^{11}| \phi_{z^{\prime}} \rangle \\					
		\end{array}
		\right)
		$}
\]
\caption{\centering The corresponding matrix $G^{zz^\prime}$, where $z = (z_0,z_1) \in \{0,1\}^2$.}
\label{Fig 3}
\end{figure*}

Then one of the code states is sent to Bob$_1$ and Bob$_2$ for selective decoding. Bob$_1$ and Bob$_2$ have the inputs $y_1,y_2 \in \{ 0,1 \}$ and the outputs $b_1,b_2 \in \{0,1\}$, respectively. The sequential measurements are labeled by $B_{y_1y_2}^{b_1b_2}$. Their goal is to guess the input bit that is associated with the position bit. For example, Bob$_1$ can choose the input $y_1 = 0$ and use his output $b_1$ to guess the value of $z_0$. The winning probabilities are defined as in Ref. \cite{Mohan_2019}:
\begin{align}
	P(b_1 =z_{y_1}) &= \frac{1}{8}\sum_{b_1 = z_{y_1}}p(b_1|y_1,z_0z_1),\notag \\
	P(b_2 =z_{y_2}) &= \frac{1}{16}\sum_{b_2 = z_{y_2}}p(b_2|y_2,z_0z_1).
	\label{eq:8}
\end{align}

 In the classical random access codes, the success guessing probability is bound by 0.75. It was shown that an optimal $2 \rightarrow 1$ QRAC for qubits can achieve a value of 0.8536, surpassing the classical bound \cite{PhysRevA.98.062307}. In the classical sequential random access codes,  a large value of $P(b_1 =z_{y_1})$ poses no obstacle to also finding a large value of $P(b_2 =z_{y_2})$. There is no trade-off between $P(b_1 =z_{y_1})$ and $P(b_2 =z_{y_2})$. The classically attainable guessing probabilities satisfy $\{P(b_i =z_{y_i})\}_{i=1,2} \in [0.5,0.75]$. In a sequential quantum model, Bob$_1$'s measurement disturbs the initial state, and therefore Bob$_2$'s ability to access the desired information depends on Bob$_1$'s preceding measurement. We consider what Bob$_2$'s optimal guessing probability is, given that Bob$_1$'s guessing probability is set to some fixed value $\tau \in [0.5, 0.8536]$. More specifically, we consider the following optimization problem:
\begin{align}
& \text{maximize:} \ && P(b_2 =z_{y_2}), \notag \\
& \text{subject to:} \ && P(b_1 =z_{y_1})=\tau, \tau \in [0.5, 0.8536]\notag \\
& && \langle \phi_z | \phi_{z^{\prime}} \rangle = \lambda_{zz^{\prime}}, \forall z, z^{\prime}, \notag\\
& && P(b_1b_2|y_1y_2,z) \in Q(\lambda)^{SEQ}.
\label{eq:9}
\end{align}

Using the adapted NPA-hierarchy, we first define the set $S_1$ as in Eq. \eqref{eq:10}, which contains all operators of the sequential receivers. 

\begin{align}
	S_1 = \{&I,B_{00}^{00},B_{00}^{01},B_{00}^{10},B_{00}^{11},B_{01}^{00},B_{01}^{01},B_{01}^{10},B_{01}^{11},\notag\\ 
	&B_{10}^{00},B_{10}^{01},B_{10}^{10},B_{10}^{11},B_{11}^{00},B_{11}^{01},B_{11}^{10},B_{11}^{11}\}.
	\label{eq:10}
\end{align}

According to the fact of four vary states $\{|\phi_{00}\rangle,|\phi_{01}\rangle,|\phi_{10}\rangle,|\phi_{11}\rangle\}$, we can partition any feasible solution $G$ in Eq. \eqref{eq:6} into 4 $\times$ 4 blocks $\{G^{zz^\prime}\}$. The corresponding $G^{zz^\prime}$ is a 17 $\times$ 17 semi-definite matrix, with elements defined as $G_{\left( i,j \right)}^{zz^{\prime}} = \langle \phi_z|{{S_1^i}}^\dagger S_1^j|\phi_{z^{\prime}}\rangle$, as illustrated in Fig. \ref{Fig 3}. Its entries satisfy certain linear relations corresponding to the algebraic constraints of the measurement operators and the inner product matrix of the prepared states. The zero entries of the matrix arise from the constraints imposed by the orthogonality of sequential measurement operators in Eq. \eqref{eq:3}. We index the rows and columns of $G^{zz^\prime}$ by \{0,1,2,...,16\}. More specifically, the matrix elements must also satisfy the following constraints:
\begin{flalign}
	 &G_{(4i-3,4i-3)}^{zz^\prime} + G_{(4i-2,4i-2)}^{zz^\prime} + G_{(4i-1,4i-1)}^{zz^\prime} + G_{(4i,4i)}^{zz^\prime} \notag \\ 
	 &= G_{(0,0)}^{zz^\prime} \notag \\ 
	 &= \lambda_{zz^{\prime}}, \quad i \in \{1,...,4\}, \notag \\ 
	& G_{(i,i)}^{zz^\prime} = G_{(0,i)}^{zz^\prime}, \quad i \in \{1,...,16\} \notag \\ 
	& G_{(i,1)}^{zz^\prime} + G_{(i,2)}^{zz^\prime} + G_{(i,3)}^{zz^\prime} + G_{(i,4)}^{zz^\prime} \notag \\
	&= G_{(i,k)}^{zz^\prime} + G_{(i,k+1)}^{zz^\prime} + G_{(i,k+2)}^{zz^\prime} + G_{(i,k+3)}^{zz^\prime}, \notag \\
	& \quad i \in \{1,...,16\}, \quad k \in \{5,9,13\}. 
	\label{eq:11}
\end{flalign}

Furthermore, Bob$_1$'s guessing probability is given by:
\begin{align}
	P(b_1 =z_{y_1}) = \frac{1}{8}\Big[&G_{(1,1)}^{00,00} + G_{(2,2)}^{00,00} + G_{(9,9)}^{00,00} + G_{(10,10)}^{00,00} \notag \\
	+ &G_{(1,1)}^{01,01} + G_{(2,2)}^{01,01} 
	+ G_{(11,11)}^{01,01} + G_{(12,12)}^{01,01} \notag \\
	+ &G_{(3,3)}^{10,10} + G_{(4,4)}^{10,10} + G_{(9,9)}^{10,10} + G_{(10,10)}^{10,10} \notag \\
	+ &G_{(3,3)}^{11,11} + G_{(4,4)}^{11,11} + G_{(11,11)}^{11,11} + G_{(12,12)}^{11,11}\Big],
	\label{Eq:12}
\end{align}
and similarly, Bob$_2$'s guessing probability is given by:
\begin{align}
	P(b_2 =z_{y_2}) = \frac{1}{16}\Big[ &G_{(1,1)}^{00,00} + G_{(3,3)}^{00,00} + G_{(5,5)}^{00,00} + G_{(7,7)}^{00,00} \notag \\
	+ &G_{(9,9)}^{00,00} + G_{(11,11)}^{00,00} + G_{(13,13)}^{00,00} + G_{(15,15)}^{00,00} \notag \\
	+ &G_{(1,1)}^{01,01} + G_{(3,3)}^{01,01} + G_{(6,6)}^{01,01} + G_{(8,8)}^{01,01} \notag \\
	+ &G_{(9,9)}^{01,01} + G_{(11,11)}^{01,01} + G_{(14,14)}^{01,01} + G_{(16,16)}^{01,01} \notag \\
	+ &G_{(2,2)}^{10,10} + G_{(4,4)}^{10,10} + G_{(5,5)}^{10,10} + G_{(7,7)}^{10,10} \notag \\
	+ &G_{(10,10)}^{10,10} + G_{(12,12)}^{10,10} + G_{(13,13)}^{10,10} + G_{(15,15)}^{10,10} \notag \\
	+ &G_{(2,2)}^{11,11} + G_{(4,4)}^{11,11} + G_{(6,6)}^{11,11} + G_{(8,8)}^{11,11} \notag \\
	+ &G_{(10,10)}^{11,11} + G_{(12,12)}^{11,11} + G_{(14,14)}^{11,11} + G_{(16,16)}^{11,11}\Big].
	\label{Eq:13}
\end{align}

\begin{figure}
	\includegraphics[width=0.5\textwidth]{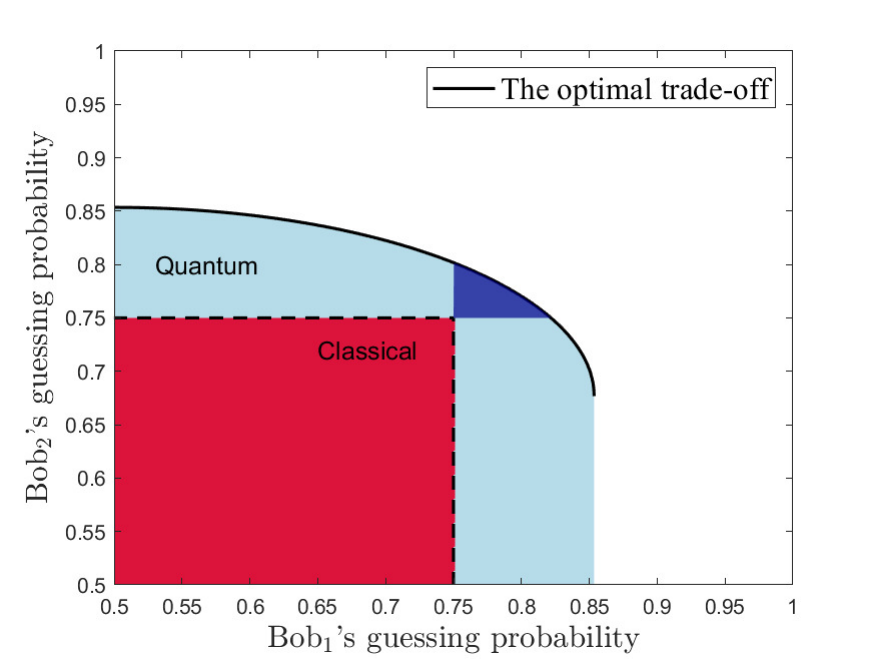}
	\caption{The correlations attainable in the space of the two guessing probabilities $(P(b_1 = z_{y_1}), P(b_2 = z_{y_2}))$ in a classical and quantum model respectively. The optimal trade-off of QRACs($\text{Bob}_1$, $\text{Bob}_2$) is highligted by a solid black line. The dark blue region indicates the simultaneous violation of the classical bounds of two guessing probabilities.} 
	\label{Fig 4} 
\end{figure}

The optimal trade-off between the pair of QRACs($\text{Bob}_1$, $\text{Bob}_2$) corresponds to $P(b_2 = z_{y_2}) = \frac{1}{8}(4 + \sqrt{2} + \sqrt{16P(b_1 = z_{y_1}) - 16{P(b_1 = z_{y_1}})^2 - 2} $ \cite{Mohan_2019}. Using the adapted NPA-hierarchy, we are able to reproduce the result (see Fig. \ref{Fig 4}). The optimal trade-off is represented by the solid black line. The red region enclosed by the black dashed line represents the area where both guessing probabilities are within the classical bound. The light blue region indicates the area where quantum implementation is possible, while the dark blue region shows where the success guessing probabilities of $\text{Bob}_1$ and $\text{Bob}_2$ simultaneously violate the classical bound of $0.75$.
 
In practice, we may opt to use a different set $S_1^\prime$ that contains fewer elements, as illustrated in Eq. (\ref{eq:15}). Every operator in the set $S_1$ can be written as a linear combination of operators in the set $S_1^\prime$. Therefore, the constraints derived from $S_1^\prime$ are at least as restrictive as those derived from $S_1$. We can partition any feasible solution $G$ in Eq. (\ref{eq:6}) into 4 $\times$ 4 blocks $\{G^{zz^\prime}\}$ , each having size 11 × 11. By reducing the dimensions of matrix $G$, the solution process becomes less resource-intensive and time-efficient.
\begin{align}
	S_1^\prime = \{&I,B_{00}^{00},B_{00}^{10},B_{01}^{00},B_{01}^{01},B_{01}^{10},B_{10}^{00},B_{10}^{10}, \notag \\
	&B_{11}^{00},B_{11}^{01},B_{11}^{10}\}.
	\label{eq:14}
\end{align}

\subsubsection{The randomness certification in the double violation region of the $2 \rightarrow 1$ sequential QRACs}
The adapted NPA-hierarchy can be used to bound the amount of randomness generated in the P\&M quantum one-chain-shaped network. The SDI randomness certification in the P\&M scenario has been studied in many previous researches \cite{PhysRevA.84.034301,PhysRevA.85.052308,PhysRevA.92.022331,Passaro_2015,PhysRevApplied.7.054018,Xiao2023}. Our approach relaxes the assumptions of the general SDI randomness certification in $2 \to 1$ QRACs \cite{PhysRevA.84.034301,PhysRevA.85.052308,PhysRevA.92.022331}, which fix the system dimensions. As mentioned above, we identify a dark blue shaded region (see Fig. \ref{Fig 4}), indicating where both Bob$_1$’s and Bob$_2$’s guessing probabilities surpass the classical threshold of 0.75. We demonstrate that within this region, both sequential receivers can successfully certify genuine randomness. To quantify the amount of randomness, we formulate an optimization problem based solely on the inner product information of the prepared states.

The guessing probability of Bob$_1$ serves as the independent variable for certifying the randomness generated by two sequential receivers. We quantify the randomness of the measurement outcomes $b_1$ and $b_2$ conditioned on the input values $y_1$, $y_2$, and $z$ by the following min-entropy function:
    \begin{align}
    H_\infty(b_1|y_1) &=  -\log_{2}{\max_{z,y_1,b_1}p(b_1|y_1,z)},\notag\\
    H_\infty(b_2|y_2) &=  -\log_{2}{\max_{z,y_2,b_2}p(b_2|y_2,z)},\notag\\
    H_\infty(b_1b_2|y_1y_2) &=  -\log_{2}{\max_{z,y_1,y_2,b_1,b_2}p(b_1b_2|y_1y_2,z)}.
    \label{eq:15}
    \end{align}  
    
More precisely, for a fixed success guessing probability of Bob$_1$, the min-entropy function $H_\infty(b_1|y_1)$ is obtained by solving the following optimization problem:
\begin{align}
	& \text{minimize:} \ && H_\infty(b_1|y_1),\notag \\
	& \text{subject to:} \ && P(b_1 =z_{y_1})=\tau, \tau \in [0.75, 0.8218], \notag\\
	& && \langle \phi_z | \phi_{z^{\prime}} \rangle = \lambda_{zz^{\prime}}, \forall z, z^{\prime}, \notag\\
	& && P(b_2 =z_{y_2}) > 0.75.
	\label{eq:16}
\end{align}
 
\begin{figure}[ht!]
	\includegraphics[width=0.5\textwidth]{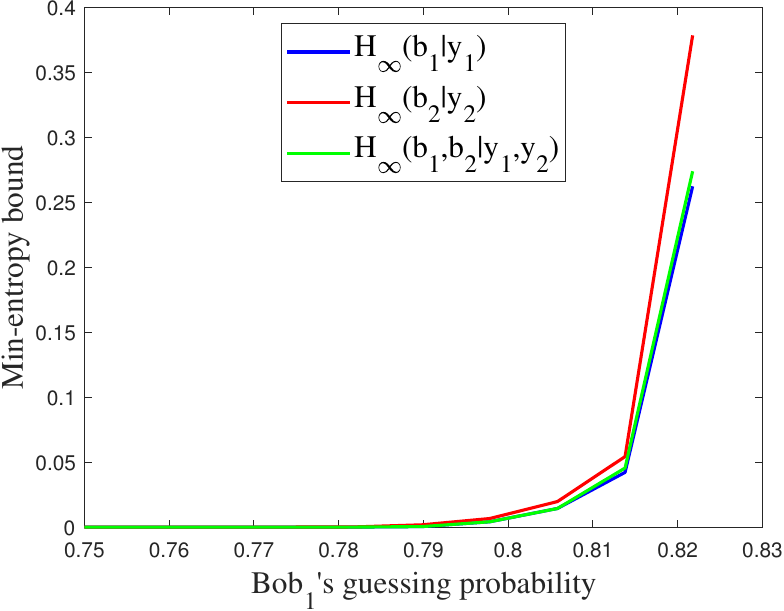}
	\caption{The min-entropy bounds $H_\infty(b_1|y_1)$, $H_\infty(b_2|y_2)$, and $H_\infty(b_1,b_2|y_1,y_2)$ presented as three functions of Bob$_1$'s guessing probability in the double violation region of $2 \to 1$ sequential QRACs.}
	\label{Fig 5}
\end{figure}

To quantify the local randomness generated by Bob$_2$, we simply convert the objective function in the optimization problem into the min-entropy $H_\infty(b_2|y_2)$. Similarly, by adjusting the objective function to $H_\infty(b_1b_2|y_1y_2)$, we quantify the global randomness of the two sequential receivers. The lower bounds of certified randomness are illustrated in Fig. \ref{Fig 5}. The amounts of local and global randomness remain zero until Bob$_1$'s guessing probability exceeds 0.79, after which they begin to increase gradually. We establish the lower bounds of randomness across the entire double violation region.
 
\subsection{The P\&M quantum two-chain-shaped network} 
\begin{figure}
	\includegraphics[width=0.5\textwidth]{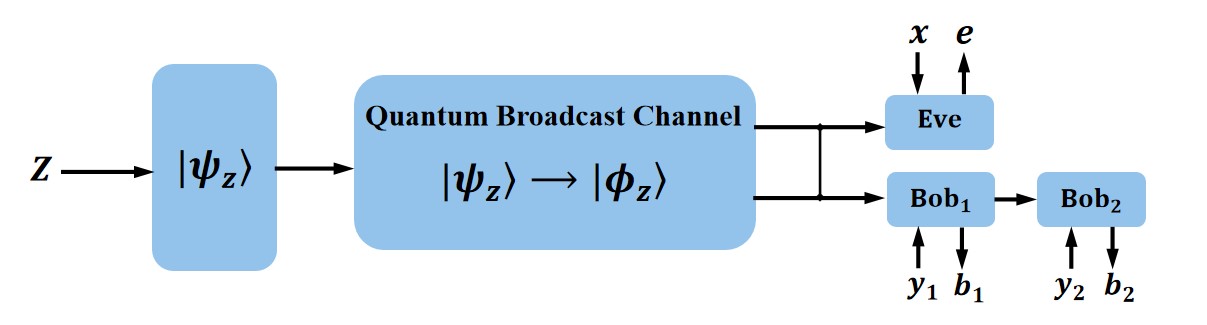}
	\caption{The P\&M quantum two-chain-shaped network with two measurement parties. One party is the eavesdropper Eve and the other party has two sequential receivers.}
	\label{Fig 6}
\end{figure}

In practice, the security of quantum cryptographic protocols is threatened by the eavesdropper Eve. The adapted NPA-hierarchy can be used to bound the amount of randomness with the presence of Eve. We use the P\&M quantum two-chain-shaped network as an example (see Fig. \ref{Fig 6}). Here, one of the parties is the eavesdropper Eve, whose goal is to guess the outcomes generated by the two sequential receivers in another measurement party. The lower bound of randomness is quantified by the min-entropy of Eve's maximum guessing probabilities.

The global guessing probability for the sequential Bob’s input $\mathbf{y}$ given a probability $P(\mathbf{b}|\mathbf{y},z)$ is the best probability that Eve could guess $\mathbf{b}$ given $ \mathbf{y} = \mathbf{y^*} $ and $z = z^*$. Simultaneously, Eve must reproduce $P(\mathbf{b}|\mathbf{y},z)$ when marginalizing over her output $e$. 
\begin{align}   
	& G(\mathbf{y},z) = && \max_{x} \sum_{e=1}^{|\mathbf{b}|} p_{BE}(\mathbf{b},e|\mathbf{y},x,z), \notag \\  
	& \text{subject to :} && P(\mathbf{b}|\mathbf{y},z) = \sum_e p_{BE}(\mathbf{b},e|\mathbf{y},x,z), \notag \\  
	& && p_{BE}(\mathbf{b},e|\mathbf{y},x,z) \in Q(\lambda)^{SEQ}. 
	\label{Eq:17}
\end{align}   

Eve has the input $x$ and the output $e \in [1,...,|\mathbf{b}|]$, where $|\mathbf{b}|$ represents the size of sequential Bob's outcomes. The second constraint represents that $p_{BE}(\mathbf{b},e|\mathbf{y},x,z)$ has a P\&M multi-chain-shaped network quantum realization satisfying the inner-product matrix $\lambda$ of the prepared states. 

The global randomness for the two sequential receivers can be quantified by the min-entropy $-\log_2G(\mathbf{y},z)$. Similarly, we can modify the objective function to calculate the local randomness of Bob$_1$ and Bob$_2$, with the constraints adjusted to focus on individual Bobs rather than the sequential Bobs.

\subsubsection{Randomness certification based on the optimal trade-off of 2→1 sequential QRACs} 
If there is an eavesdropper Eve as shown in Fig. \ref{Fig 6}, what amount of local and global randomness can be certified in 2→1 sequential QRACs? To better capture the quantum characteristics, we can constrain the guessing probabilities of Bob$_1$ and Bob$_2$ to satisfy the optimal trade-off relationship, rather than solving across the double violation quantum region. That is:
\begin{align}   
	& G(\mathbf{y},z) = && \max_{x} \sum_{e=1}^{|\mathbf{b}|} p_{BE}(\mathbf{b},e|\mathbf{y},x,z), \notag \\  
	& \text{subject to :} &&  P(b_1 =z_{y_1})=\tau, \tau \in [0.5, 0.8536],\notag \\  
	& &&P(b_2 = z_{y_2}) = \frac{1}{8}(4 + \sqrt{2} +  \notag \\ 
	& &&\sqrt{16P(b_1 = z_{y_1}) - 16{P(b_1 = z_{y_1}})^2 - 2},  \notag \\ 
	& && \sum_ep_{BE}(\mathbf{b},e|\mathbf{y},x,z) = p(\mathbf{b},|\mathbf{y},z), \forall y,z. 
	\label{Eq:18}
\end{align} 

\begin{figure}
	\includegraphics[width=0.5\textwidth]{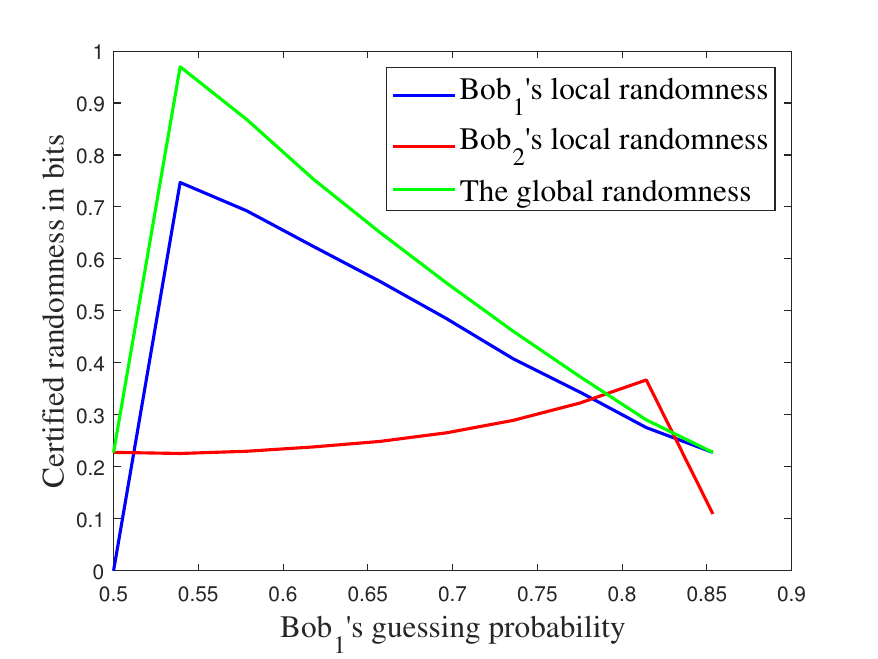}
	\caption{The relationships between the randomness generated by two sequential receivers and Bob$_1$'s guessing probability. The guessing probabilities of Bob$_1$ and Bob$_2$ satisfy the optimal trade-off relationship.}
	\label{Fig 7}
\end{figure}

Each measurement choice for Bob$_1$ and Bob$_2$ yields two possible outcomes, resulting in four joint outcomes $b_1b_2$. Eve aims to guess the joint outcome. It is assumed that Eve has a single input and four outputs. Bob$_1$'s guessing probability is used as the variable, while Bob$_2$'s guessing probability is optimized. The final constraint ensures that the sum of Eve's outcomes matches the joint probability in the scenario without Eve's involvement. When certifying local randomness of Bob$_1$ and Bob$_2$, it suffices to assume that Eve has two possible outcomes. In the optimization problem in Eq. \eqref{Eq:18}, the objective function should be adjusted to the local guessing probability. The certified randomness obtained using the adapted NPA-hierarchy is presented in the Fig. \ref{Fig 7}.

 \subsubsection{Randomness certification based on complete set of probabilities}
In the previous section, randomness is certified based on the optimal trade-off of $2 \rightarrow 1$ sequential QRACs, where the guessing probabilities align with dimension witness inequality \cite{PhysRevA.85.052308}. However, a complete set of probabilities always contains more information than a single inequality. Nieto-Silleras et al. have already utilized a complete set of probabilities for optimal device-independent randomness evaluation \cite{Nieto-Silleras_2014}. Here, we will investigate the SDI randomness certification from the complete probability distribution $P_{obs}$ in the presence of an eavesdropper. The $P_{obs}$ is generated by two sequential receivers, Bob$_1$ and Bob$_2$. 

Our SDI randomness certification protocol uses the BB84 states $\{|0\rangle,|+\rangle,|-\rangle,|1\rangle \}$. One of these four states is randomly distributed to Bob$_1$, who performs one of two unsharp measurements corresponding to $\eta \sigma_x $ and $\eta \sigma_z $ for a sharpness parameter $\eta \in [0,1]$. Bob$_1$ then sends his post-measurement state to Bob$_2$. For $y_2 = 0,1$, Bob$_2$ performs a projective measurement of $\sigma_x$ or $\sigma_z$. The eavesdropper Eve performs measurement $x$ to guess the global outcome $b_1b_2$ or the local outcomes $b_1$ and $b_2$. The optimization problem for certifying global randomness is formulated as follows: 
\begin{align}   
	& G(\mathbf{y},z) = && \max_{x} \sum_{e=1}^{|\mathbf{b}|} p_{BE}(\mathbf{b},e|\mathbf{y},x,z), \notag \\  
	& \text{subject to :} && P_{obs}(\mathbf{b}|\mathbf{y},z) = \sum_e p_{BE}(\mathbf{b},e|\mathbf{y},x,z), \notag \\  
	& && p_{BE}(\mathbf{b},e|\mathbf{y},x,z) \in Q(\lambda)^{SEQ}. 
	\label{Eq:19}
\end{align} 
 \begin{figure}
	\includegraphics[width=0.5\textwidth]{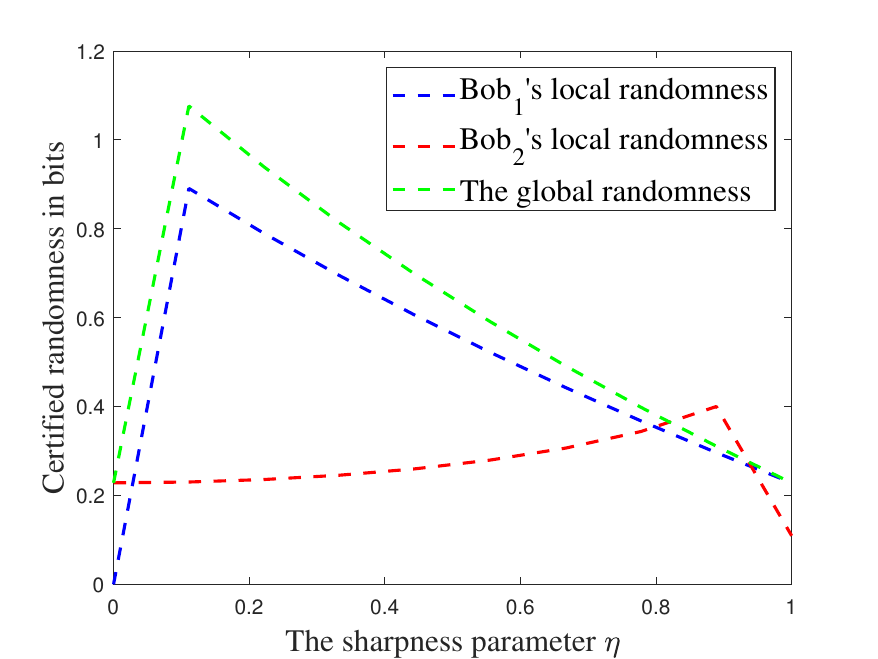}
	\caption{The relationships between the randomness generated by the two sequential receivers and Bob$_1$'s sharpness parameter $\eta$.} 
	\label{Fig 8} 
\end{figure}
\begin{figure*}[ht!] 
	\centering
	\begin{subfigure}[b]{0.48\textwidth} 
		\centering
		\includegraphics[width=\textwidth]{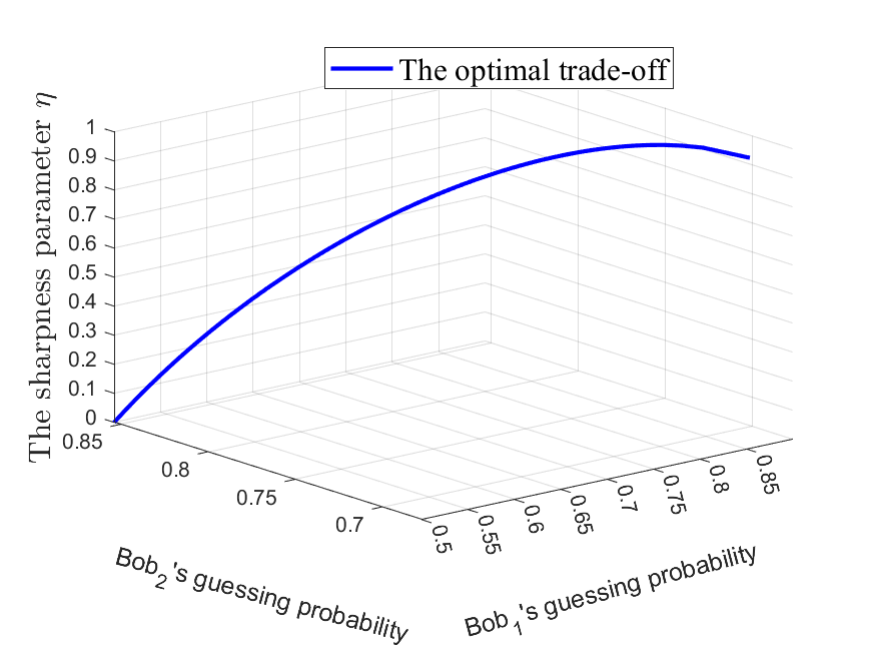} 
		\caption{\centering}
		\label{fig:subfigure1}
	\end{subfigure}
	\hfill
	\begin{subfigure}[b]{0.48\textwidth} 
		\centering
		\includegraphics[width=\textwidth]{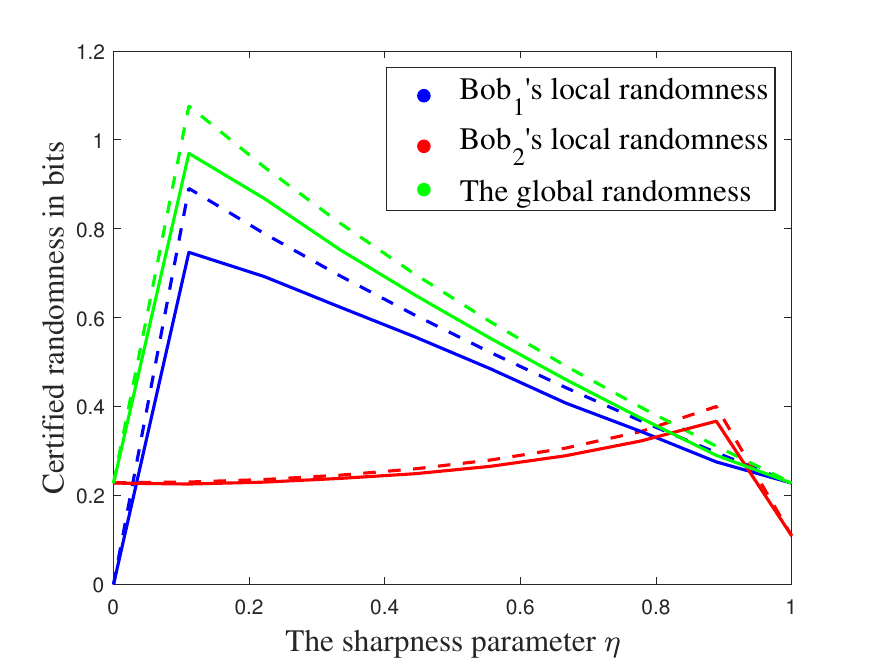} 
		\caption{\centering}
		\label{fig:subfigure2}
	\end{subfigure}
	\caption{(a). The trade-off between the guessing probabilities of Bob$_1$ and Bob$_2$ as the sharpness parameter ranges from 0 to 1. (b). The certified randomness as a function of the sharpness parameter $\eta$. The solid curve is obtained by taking into account only the optimal trade-off relationship, while the dashed
		curve is obtained by taking into account the complete set of probabilities.} 
	\label{Fig 9} 
\end{figure*}
The first constraint requires that the sum of Eve's outcomes matches the observed probability values. The second constraint ensures that the probability distribution produced by Eve and the sequential Bobs lies within the quantum set. By adjusting the objective function and the number of outcomes generated by Eve, the amount of local randomness can be certified. Since Bob$_1$'s sharpness measurement parameter influences the amount of information obtained by both Bob$_1$ and Bob$_2$, we provide the relationship between the sharpness parameter $\eta$ and the certified random bits. In Fig. \ref{Fig 8} we present the min-entropy $-\log_2(G({y_1y_2}^* = 00, z^* = 1))$, $-\log_2(G({y_1}^* = 0, z^* = 1))$ and $-\log_2(G({y_2}^* = 0, z^* = 1))$ obtained as functions of the sharpness parameter $\eta$.
\setlength{\parskip}{2pt}

From Fig. \ref{Fig 8}, we can observe that the trend of local or global randomness as a function of the sharpness parameter aligns with that shown in Fig. \ref{Fig 7}. The complete set of probabilities is generated from four BB84 states, which can be used to implement $2 \rightarrow 1$ sequential QRACs. As the sharpness parameter $\eta$ varies from 0 to 1, Bob$_1$'s success guessing probability increases from 0.5 to 0.8536. During this range, the relationship between the guessing probabilities of Bob$_1$ and Bob$_2$ aligns with the optimal trade-off in $2 \rightarrow 1$ sequential QRACs (see Fig. \ref{Fig 9}(a)). Thus, by employing the sharpness parameter as an independent variable, we can compare the certified randomness derived from the optimal trade-off with that obtained from the complete probabilities, concluding that the latter certifies more randomness (see Fig. \ref{Fig 9}(b)). This result aligns with the assertion in Ref. \cite{Bancal_2014}, which claims that more randomness can be certified when the full set of probabilities is considered.
\section{Conclusion}	
The approach introduced in this work provides a comprehensive toolbox for characterizing correlations generated in P\&M quantum multi-chain-shaped networks, featuring sequential receivers at each measurement party. Assuming only the inner product information of prepared states, we derive a hierarchy of necessary conditions for the correlations arising from P\&M quantum multi-chain-shaped networks. Each necessary condition amounts to verifyng the existence of a positive semi-definite matrix $G$. It satisfies the linear and positive semi-definite constraints generated by the inner-product of prepared states and sequential measurement operators.

Optimizing the entire quantum correlation set has many practical applications, one of which is the sequential QRACs. Using our adapted NPA-hierarchy, we derived the optimal trade-off between the success guessing probabilities of two sequential receivers in $2 \rightarrow 1$ sequential QRACs. In principle, the optimal trade-off obtained from the low-level NPA-hierarchy may not be tight. This is because the hierarchy only provides necessary conditions for determining whether the given correlations are generated in the P\&M quantum multi-chain-shaped networks. In the application, we show that low-level approximations are already enough to achieve tight bounds. Furthermore, the adapted NPA-hierarchy can also be used for SDI randomness certification in the P\&M quantum multi-chain-shaped networks. We first examine randomness certification within the double violation region of $2 \rightarrow 1$ sequential QRACs. Then considering the presence of an eavesdropper Eve in practical communication, we investigate the randomness that can be certified using both the optimal trade-off of $2 \to 1$ sequential QRACs and the complete probabilities generated by two sequential receivers. We find that utilizing the complete probabilities yields more certified randomness than relying solely on the optimal trade-off.

Given the advantages of quantum correlation sets, it is worth exploring whether our toolbox can be applied to quantum information tasks beyond those discussed in this paper. The tasks addressed here focus on simple P\&M network scenarios with a limited number of measurement parties. However, as we extend our attention to P\&M quantum multi-chain networks that involve a greater number of measurement parties, the time and memory requirements of the NPA-hierarchy become increasingly prohibitive. Machine learning methods can effectively characterize quantum correlations within quantum systems while also minimizing resource consumption \cite{PhysRevLett.122.200401, Song2024}. For instance, the approach outlined in Ref. \cite{PhysRevResearch.5.023016} utilizes a feedforward artificial neural network to optimize correlations compatible with arbitrary quantum networks. We could consider adopting a similar strategy to characterize the correlations generated in more complex P\&M networks.

\section*{Acknowledgment}
This work is supported by the National Natural Science Foundation of China (Grants No.62171056, No.62101600, and No.62220106012).


\begin{thebibliography}{52}%
	\makeatletter
	\providecommand \@ifxundefined [1]{%
		\@ifx{#1\undefined}
	}%
	\providecommand \@ifnum [1]{%
		\ifnum #1\expandafter \@firstoftwo
		\else \expandafter \@secondoftwo
		\fi
	}%
	\providecommand \@ifx [1]{%
		\ifx #1\expandafter \@firstoftwo
		\else \expandafter \@secondoftwo
		\fi
	}%
	\providecommand \natexlab [1]{#1}%
	\providecommand \enquote  [1]{``#1''}%
	\providecommand \bibnamefont  [1]{#1}%
	\providecommand \bibfnamefont [1]{#1}%
	\providecommand \citenamefont [1]{#1}%
	\providecommand \href@noop [0]{\@secondoftwo}%
	\providecommand \href [0]{\begingroup \@sanitize@url \@href}%
	\providecommand \@href[1]{\@@startlink{#1}\@@href}%
	\providecommand \@@href[1]{\endgroup#1\@@endlink}%
	\providecommand \@sanitize@url [0]{\catcode `\\12\catcode `\$12\catcode
		`\&12\catcode `\#12\catcode `\^12\catcode `\_12\catcode `\%12\relax}%
	\providecommand \@@startlink[1]{}%
	\providecommand \@@endlink[0]{}%
	\providecommand \url  [0]{\begingroup\@sanitize@url \@url }%
	\providecommand \@url [1]{\endgroup\@href {#1}{\urlprefix }}%
	\providecommand \urlprefix  [0]{URL }%
	\providecommand \Eprint [0]{\href }%
	\providecommand \doibase [0]{https://doi.org/}%
	\providecommand \selectlanguage [0]{\@gobble}%
	\providecommand \bibinfo  [0]{\@secondoftwo}%
	\providecommand \bibfield  [0]{\@secondoftwo}%
	\providecommand \translation [1]{[#1]}%
	\providecommand \BibitemOpen [0]{}%
	\providecommand \bibitemStop [0]{}%
	\providecommand \bibitemNoStop [0]{.\EOS\space}%
	\providecommand \EOS [0]{\spacefactor3000\relax}%
	\providecommand \BibitemShut  [1]{\csname bibitem#1\endcsname}%
	\let\auto@bib@innerbib\@empty
	\bibitem [{\citenamefont {Horodecki}\ \emph {et~al.}(2009)\citenamefont
		{Horodecki}, \citenamefont {Horodecki}, \citenamefont {Horodecki},\ and\
		\citenamefont {Horodecki}}]{RevModPhys.81.865}%
	\BibitemOpen
	\bibfield  {author} {\bibinfo {author} {\bibfnamefont {R.}~\bibnamefont
			{Horodecki}}, \bibinfo {author} {\bibfnamefont {P.}~\bibnamefont
			{Horodecki}}, \bibinfo {author} {\bibfnamefont {M.}~\bibnamefont
			{Horodecki}},\ and\ \bibinfo {author} {\bibfnamefont {K.}~\bibnamefont
			{Horodecki}},\ }\bibfield  {title} {\bibinfo {title} {Quantum entanglement},\
	}\href {https://doi.org/10.1103/RevModPhys.81.865} {\bibfield  {journal}
		{\bibinfo  {journal} {Rev. Mod. Phys.}\ }\textbf {\bibinfo {volume} {81}},\
		\bibinfo {pages} {865} (\bibinfo {year} {2009})}\BibitemShut {NoStop}%
	\bibitem [{\citenamefont {Wiseman}\ \emph {et~al.}(2007)\citenamefont
		{Wiseman}, \citenamefont {Jones},\ and\ \citenamefont
		{Doherty}}]{PhysRevLett.98.140402}%
	\BibitemOpen
	\bibfield  {author} {\bibinfo {author} {\bibfnamefont {H.~M.}\ \bibnamefont
			{Wiseman}}, \bibinfo {author} {\bibfnamefont {S.~J.}\ \bibnamefont {Jones}},\
		and\ \bibinfo {author} {\bibfnamefont {A.~C.}\ \bibnamefont {Doherty}},\
	}\bibfield  {title} {\bibinfo {title} {Steering, entanglement, nonlocality,
			and the einstein-podolsky-rosen paradox},\ }\href
	{https://doi.org/10.1103/PhysRevLett.98.140402} {\bibfield  {journal}
		{\bibinfo  {journal} {Phys. Rev. Lett.}\ }\textbf {\bibinfo {volume} {98}},\
		\bibinfo {pages} {140402} (\bibinfo {year} {2007})}\BibitemShut {NoStop}%
	\bibitem [{\citenamefont {Brunner}\ \emph {et~al.}(2014)\citenamefont
		{Brunner}, \citenamefont {Cavalcanti}, \citenamefont {Pironio}, \citenamefont
		{Scarani},\ and\ \citenamefont {Wehner}}]{RevModPhys.86.419}%
	\BibitemOpen
	\bibfield  {author} {\bibinfo {author} {\bibfnamefont {N.}~\bibnamefont
			{Brunner}}, \bibinfo {author} {\bibfnamefont {D.}~\bibnamefont {Cavalcanti}},
		\bibinfo {author} {\bibfnamefont {S.}~\bibnamefont {Pironio}}, \bibinfo
		{author} {\bibfnamefont {V.}~\bibnamefont {Scarani}},\ and\ \bibinfo {author}
		{\bibfnamefont {S.}~\bibnamefont {Wehner}},\ }\bibfield  {title} {\bibinfo
		{title} {Bell nonlocality},\ }\href
	{https://doi.org/10.1103/RevModPhys.86.419} {\bibfield  {journal} {\bibinfo
			{journal} {Rev. Mod. Phys.}\ }\textbf {\bibinfo {volume} {86}},\ \bibinfo
		{pages} {419} (\bibinfo {year} {2014})}\BibitemShut {NoStop}%
	\bibitem [{\citenamefont {Cavalcanti}\ \emph {et~al.}(2011)\citenamefont
		{Cavalcanti}, \citenamefont {Almeida}, \citenamefont {Scarani},\ and\
		\citenamefont {Acín}}]{Cavalcanti_2011}%
	\BibitemOpen
	\bibfield  {author} {\bibinfo {author} {\bibfnamefont {D.}~\bibnamefont
			{Cavalcanti}}, \bibinfo {author} {\bibfnamefont {M.~L.}\ \bibnamefont
			{Almeida}}, \bibinfo {author} {\bibfnamefont {V.}~\bibnamefont {Scarani}},\
		and\ \bibinfo {author} {\bibfnamefont {A.}~\bibnamefont {Acín}},\ }\bibfield
	{title} {\bibinfo {title} {Quantum networks reveal quantum nonlocality},\
	}\bibfield  {journal} {\bibinfo  {journal} {Nature Communications}\ }\textbf
	{\bibinfo {volume} {2}},\ \href {https://doi.org/10.1038/ncomms1193}
	{10.1038/ncomms1193} (\bibinfo {year} {2011})\BibitemShut {NoStop}%
	\bibitem [{\citenamefont {Spekkens}(2005)}]{PhysRevA.71.052108}%
	\BibitemOpen
	\bibfield  {author} {\bibinfo {author} {\bibfnamefont {R.~W.}\ \bibnamefont
			{Spekkens}},\ }\bibfield  {title} {\bibinfo {title} {Contextuality for
			preparations, transformations, and unsharp measurements},\ }\href
	{https://doi.org/10.1103/PhysRevA.71.052108} {\bibfield  {journal} {\bibinfo
			{journal} {Phys. Rev. A}\ }\textbf {\bibinfo {volume} {71}},\ \bibinfo
		{pages} {052108} (\bibinfo {year} {2005})}\BibitemShut {NoStop}%
	\bibitem [{\citenamefont {Ac\'{\i}n}\ \emph {et~al.}(2012)\citenamefont
		{Ac\'{\i}n}, \citenamefont {Massar},\ and\ \citenamefont
		{Pironio}}]{PhysRevLett.108.100402}%
	\BibitemOpen
	\bibfield  {author} {\bibinfo {author} {\bibfnamefont {A.}~\bibnamefont
			{Ac\'{\i}n}}, \bibinfo {author} {\bibfnamefont {S.}~\bibnamefont {Massar}},\
		and\ \bibinfo {author} {\bibfnamefont {S.}~\bibnamefont {Pironio}},\
	}\bibfield  {title} {\bibinfo {title} {Randomness versus nonlocality and
			entanglement},\ }\href {https://doi.org/10.1103/PhysRevLett.108.100402}
	{\bibfield  {journal} {\bibinfo  {journal} {Phys. Rev. Lett.}\ }\textbf
		{\bibinfo {volume} {108}},\ \bibinfo {pages} {100402} (\bibinfo {year}
		{2012})}\BibitemShut {NoStop}%
	\bibitem [{\citenamefont {Dhara}\ \emph {et~al.}(2013)\citenamefont {Dhara},
		\citenamefont {Prettico},\ and\ \citenamefont
		{Ac\'{\i}n}}]{PhysRevA.88.052116}%
	\BibitemOpen
	\bibfield  {author} {\bibinfo {author} {\bibfnamefont {C.}~\bibnamefont
			{Dhara}}, \bibinfo {author} {\bibfnamefont {G.}~\bibnamefont {Prettico}},\
		and\ \bibinfo {author} {\bibfnamefont {A.}~\bibnamefont {Ac\'{\i}n}},\
	}\bibfield  {title} {\bibinfo {title} {Maximal quantum randomness in bell
			tests},\ }\href {https://doi.org/10.1103/PhysRevA.88.052116} {\bibfield
		{journal} {\bibinfo  {journal} {Phys. Rev. A}\ }\textbf {\bibinfo {volume}
			{88}},\ \bibinfo {pages} {052116} (\bibinfo {year} {2013})}\BibitemShut
	{NoStop}%
	\bibitem [{\citenamefont {Harumichi~NISHIMURA}(2009)}]{e92-a_5_1268}%
	\BibitemOpen
	\bibfield  {author} {\bibinfo {author} {\bibfnamefont {R.~R.}\ \bibnamefont
			{Harumichi~NISHIMURA}},\ }\bibfield  {title} {\bibinfo {title} {Quantum
			random access coding},\ }\href {https://doi.org/10.1587/transfun.E92.A.1268}
	{\bibfield  {journal} {\bibinfo  {journal} {IEICE TRANSACTIONS on
				Fundamentals}\ }\textbf {\bibinfo {volume} {E92-A}},\ \bibinfo {pages} {1268}
		(\bibinfo {year} {2009})}\BibitemShut {NoStop}%
	\bibitem [{\citenamefont {Navascu\'es}\ \emph {et~al.}(2007)\citenamefont
		{Navascu\'es}, \citenamefont {Pironio},\ and\ \citenamefont
		{Ac\'{\i}n}}]{PhysRevLett.98.010401}%
	\BibitemOpen
	\bibfield  {author} {\bibinfo {author} {\bibfnamefont {M.}~\bibnamefont
			{Navascu\'es}}, \bibinfo {author} {\bibfnamefont {S.}~\bibnamefont
			{Pironio}},\ and\ \bibinfo {author} {\bibfnamefont {A.}~\bibnamefont
			{Ac\'{\i}n}},\ }\bibfield  {title} {\bibinfo {title} {Bounding the set of
			quantum correlations},\ }\href
	{https://doi.org/10.1103/PhysRevLett.98.010401} {\bibfield  {journal}
		{\bibinfo  {journal} {Phys. Rev. Lett.}\ }\textbf {\bibinfo {volume} {98}},\
		\bibinfo {pages} {010401} (\bibinfo {year} {2007})}\BibitemShut {NoStop}%
	\bibitem [{\citenamefont {Navascués}\ \emph {et~al.}(2008)\citenamefont
		{Navascués}, \citenamefont {Pironio},\ and\ \citenamefont
		{Acín}}]{Navascués_2008}%
	\BibitemOpen
	\bibfield  {author} {\bibinfo {author} {\bibfnamefont {M.}~\bibnamefont
			{Navascués}}, \bibinfo {author} {\bibfnamefont {S.}~\bibnamefont
			{Pironio}},\ and\ \bibinfo {author} {\bibfnamefont {A.}~\bibnamefont
			{Acín}},\ }\bibfield  {title} {\bibinfo {title} {A convergent hierarchy of
			semidefinite programs characterizing the set of quantum correlations},\
	}\href {https://doi.org/10.1088/1367-2630/10/7/073013} {\bibfield  {journal}
		{\bibinfo  {journal} {New J. Phys.}\ }\textbf {\bibinfo {volume} {10}},\
		\bibinfo {pages} {073013} (\bibinfo {year} {2008})}\BibitemShut {NoStop}%
	\bibitem [{\citenamefont {Fritz}(2012)}]{Fritz_2012}%
	\BibitemOpen
	\bibfield  {author} {\bibinfo {author} {\bibfnamefont {T.}~\bibnamefont
			{Fritz}},\ }\bibfield  {title} {\bibinfo {title} {Beyond bell's theorem:
			correlation scenarios},\ }\href
	{https://doi.org/10.1088/1367-2630/14/10/103001} {\bibfield  {journal}
		{\bibinfo  {journal} {New Journal of Physics}\ }\textbf {\bibinfo {volume}
			{14}},\ \bibinfo {pages} {103001} (\bibinfo {year} {2012})}\BibitemShut
	{NoStop}%
	\bibitem [{\citenamefont {Poderini}\ \emph {et~al.}(2021)\citenamefont
		{Poderini}, \citenamefont {Agresti}, \citenamefont {Marchese}, \citenamefont
		{Polino}, \citenamefont {Giordani}, \citenamefont {Suprano}, \citenamefont
		{Valeri}, \citenamefont {Milani}, \citenamefont {Spagnolo}, \citenamefont
		{Carvacho}, \citenamefont {Chaves},\ and\ \citenamefont
		{Sciarrino}}]{9571334}%
	\BibitemOpen
	\bibfield  {author} {\bibinfo {author} {\bibfnamefont {D.}~\bibnamefont
			{Poderini}}, \bibinfo {author} {\bibfnamefont {I.}~\bibnamefont {Agresti}},
		\bibinfo {author} {\bibfnamefont {G.}~\bibnamefont {Marchese}}, \bibinfo
		{author} {\bibfnamefont {E.}~\bibnamefont {Polino}}, \bibinfo {author}
		{\bibfnamefont {T.}~\bibnamefont {Giordani}}, \bibinfo {author}
		{\bibfnamefont {A.}~\bibnamefont {Suprano}}, \bibinfo {author} {\bibfnamefont
			{M.}~\bibnamefont {Valeri}}, \bibinfo {author} {\bibfnamefont
			{G.}~\bibnamefont {Milani}}, \bibinfo {author} {\bibfnamefont
			{N.}~\bibnamefont {Spagnolo}}, \bibinfo {author} {\bibfnamefont
			{G.}~\bibnamefont {Carvacho}}, \bibinfo {author} {\bibfnamefont
			{R.}~\bibnamefont {Chaves}},\ and\ \bibinfo {author} {\bibfnamefont
			{F.}~\bibnamefont {Sciarrino}},\ }\bibfield  {title} {\bibinfo {title}
		{Experimental violation of n-locality in a star quantum network[1]},\ }in\
	\href@noop {} {\emph {\bibinfo {booktitle} {2021 Conference on Lasers and
				Electro-Optics (CLEO)}}}\ (\bibinfo {year} {2021})\ pp.\ \bibinfo {pages}
	{1--3}\BibitemShut {NoStop}%
	\bibitem [{\citenamefont {Andreoli}\ \emph {et~al.}(2017)\citenamefont
		{Andreoli}, \citenamefont {Carvacho}, \citenamefont {Santodonato},
		\citenamefont {Chaves},\ and\ \citenamefont {Sciarrino}}]{Andreoli_2017}%
	\BibitemOpen
	\bibfield  {author} {\bibinfo {author} {\bibfnamefont {F.}~\bibnamefont
			{Andreoli}}, \bibinfo {author} {\bibfnamefont {G.}~\bibnamefont {Carvacho}},
		\bibinfo {author} {\bibfnamefont {L.}~\bibnamefont {Santodonato}}, \bibinfo
		{author} {\bibfnamefont {R.}~\bibnamefont {Chaves}},\ and\ \bibinfo {author}
		{\bibfnamefont {F.}~\bibnamefont {Sciarrino}},\ }\bibfield  {title} {\bibinfo
		{title} {Maximal qubit violation of n-locality inequalities in a star-shaped
			quantum network},\ }\href {https://doi.org/10.1088/1367-2630/aa8b9b}
	{\bibfield  {journal} {\bibinfo  {journal} {New J. Phys.}\ }\textbf {\bibinfo
			{volume} {19}},\ \bibinfo {pages} {113020} (\bibinfo {year}
		{2017})}\BibitemShut {NoStop}%
	\bibitem [{\citenamefont {Munshi}\ \emph {et~al.}(2021)\citenamefont {Munshi},
		\citenamefont {Kumar},\ and\ \citenamefont {Pan}}]{PhysRevA.104.042217}%
	\BibitemOpen
	\bibfield  {author} {\bibinfo {author} {\bibfnamefont {S.}~\bibnamefont
			{Munshi}}, \bibinfo {author} {\bibfnamefont {R.}~\bibnamefont {Kumar}},\ and\
		\bibinfo {author} {\bibfnamefont {A.~K.}\ \bibnamefont {Pan}},\ }\bibfield
	{title} {\bibinfo {title} {Generalized $n$-locality inequalities in a
			star-network configuration and their optimal quantum violations},\ }\href
	{https://doi.org/10.1103/PhysRevA.104.042217} {\bibfield  {journal} {\bibinfo
			{journal} {Phys. Rev. A}\ }\textbf {\bibinfo {volume} {104}},\ \bibinfo
		{pages} {042217} (\bibinfo {year} {2021})}\BibitemShut {NoStop}%
	\bibitem [{\citenamefont {Tavakoli}\ \emph {et~al.}(2014)\citenamefont
		{Tavakoli}, \citenamefont {Skrzypczyk}, \citenamefont {Cavalcanti},\ and\
		\citenamefont {Ac\'{\i}n}}]{PhysRevA.90.062109}%
	\BibitemOpen
	\bibfield  {author} {\bibinfo {author} {\bibfnamefont {A.}~\bibnamefont
			{Tavakoli}}, \bibinfo {author} {\bibfnamefont {P.}~\bibnamefont
			{Skrzypczyk}}, \bibinfo {author} {\bibfnamefont {D.}~\bibnamefont
			{Cavalcanti}},\ and\ \bibinfo {author} {\bibfnamefont {A.}~\bibnamefont
			{Ac\'{\i}n}},\ }\bibfield  {title} {\bibinfo {title} {Nonlocal correlations
			in the star-network configuration},\ }\href
	{https://doi.org/10.1103/PhysRevA.90.062109} {\bibfield  {journal} {\bibinfo
			{journal} {Phys. Rev. A}\ }\textbf {\bibinfo {volume} {90}},\ \bibinfo
		{pages} {062109} (\bibinfo {year} {2014})}\BibitemShut {NoStop}%
	\bibitem [{\citenamefont {Yang}\ \emph {et~al.}(2023)\citenamefont {Yang},
		\citenamefont {He}, \citenamefont {Qi},\ and\ \citenamefont
		{Hou}}]{Yang:2023ylc}%
	\BibitemOpen
	\bibfield  {author} {\bibinfo {author} {\bibfnamefont {S.}~\bibnamefont
			{Yang}}, \bibinfo {author} {\bibfnamefont {K.}~\bibnamefont {He}}, \bibinfo
		{author} {\bibfnamefont {X.}~\bibnamefont {Qi}},\ and\ \bibinfo {author}
		{\bibfnamefont {J.}~\bibnamefont {Hou}},\ }\bibfield  {title} {\bibinfo
		{title} {{Quantum steering in two-forked tree-shaped networks}},\ }\href
	{https://doi.org/10.1088/1402-4896/ad049f} {\bibfield  {journal} {\bibinfo
			{journal} {Phys. Scripta}\ }\textbf {\bibinfo {volume} {98}},\ \bibinfo
		{pages} {125102} (\bibinfo {year} {2023})}\BibitemShut {NoStop}%
	\bibitem [{\citenamefont {Sun}\ \emph {et~al.}(2024)\citenamefont {Sun},
		\citenamefont {Guo}, \citenamefont {Dong},\ and\ \citenamefont
		{Gao}}]{PhysRevA.110.012401}%
	\BibitemOpen
	\bibfield  {author} {\bibinfo {author} {\bibfnamefont {H.}~\bibnamefont
			{Sun}}, \bibinfo {author} {\bibfnamefont {F.}~\bibnamefont {Guo}}, \bibinfo
		{author} {\bibfnamefont {H.}~\bibnamefont {Dong}},\ and\ \bibinfo {author}
		{\bibfnamefont {F.}~\bibnamefont {Gao}},\ }\bibfield  {title} {\bibinfo
		{title} {Network nonlocality sharing in a two-forked tree-shaped network},\
	}\href {https://doi.org/10.1103/PhysRevA.110.012401} {\bibfield  {journal}
		{\bibinfo  {journal} {Phys. Rev. A}\ }\textbf {\bibinfo {volume} {110}},\
		\bibinfo {pages} {012401} (\bibinfo {year} {2024})}\BibitemShut {NoStop}%
	\bibitem [{\citenamefont {Gallego}\ \emph {et~al.}(2014)\citenamefont
		{Gallego}, \citenamefont {Würflinger}, \citenamefont {Chaves}, \citenamefont
		{Acín},\ and\ \citenamefont {Navascués}}]{Gallego_2014}%
	\BibitemOpen
	\bibfield  {author} {\bibinfo {author} {\bibfnamefont {R.}~\bibnamefont
			{Gallego}}, \bibinfo {author} {\bibfnamefont {L.~E.}\ \bibnamefont
			{Würflinger}}, \bibinfo {author} {\bibfnamefont {R.}~\bibnamefont {Chaves}},
		\bibinfo {author} {\bibfnamefont {A.}~\bibnamefont {Acín}},\ and\ \bibinfo
		{author} {\bibfnamefont {M.}~\bibnamefont {Navascués}},\ }\bibfield  {title}
	{\bibinfo {title} {Nonlocality in sequential correlation scenarios},\ }\href
	{https://doi.org/10.1088/1367-2630/16/3/033037} {\bibfield  {journal}
		{\bibinfo  {journal} {New J. Phys.}\ }\textbf {\bibinfo {volume} {16}},\
		\bibinfo {pages} {033037} (\bibinfo {year} {2014})}\BibitemShut {NoStop}%
	\bibitem [{\citenamefont {Doolittle}\ and\ \citenamefont
		{Chitambar}(2023)}]{PhysRevA.108.042409}%
	\BibitemOpen
	\bibfield  {author} {\bibinfo {author} {\bibfnamefont {B.}~\bibnamefont
			{Doolittle}}\ and\ \bibinfo {author} {\bibfnamefont {E.}~\bibnamefont
			{Chitambar}},\ }\bibfield  {title} {\bibinfo {title} {Maximal qubit
			violations of $n$-locality in star and chain networks},\ }\href
	{https://doi.org/10.1103/PhysRevA.108.042409} {\bibfield  {journal} {\bibinfo
			{journal} {Phys. Rev. A}\ }\textbf {\bibinfo {volume} {108}},\ \bibinfo
		{pages} {042409} (\bibinfo {year} {2023})}\BibitemShut {NoStop}%
	\bibitem [{\citenamefont {Pozas-Kerstjens}\ \emph {et~al.}(2019)\citenamefont
		{Pozas-Kerstjens}, \citenamefont {Rabelo}, \citenamefont {Rudnicki},
		\citenamefont {Chaves}, \citenamefont {Cavalcanti}, \citenamefont
		{Navascu\'es},\ and\ \citenamefont {Ac\'{\i}n}}]{PhysRevLett.123.140503}%
	\BibitemOpen
	\bibfield  {author} {\bibinfo {author} {\bibfnamefont {A.}~\bibnamefont
			{Pozas-Kerstjens}}, \bibinfo {author} {\bibfnamefont {R.}~\bibnamefont
			{Rabelo}}, \bibinfo {author} {\bibfnamefont {L.}~\bibnamefont {Rudnicki}},
		\bibinfo {author} {\bibfnamefont {R.}~\bibnamefont {Chaves}}, \bibinfo
		{author} {\bibfnamefont {D.}~\bibnamefont {Cavalcanti}}, \bibinfo {author}
		{\bibfnamefont {M.}~\bibnamefont {Navascu\'es}},\ and\ \bibinfo {author}
		{\bibfnamefont {A.}~\bibnamefont {Ac\'{\i}n}},\ }\bibfield  {title} {\bibinfo
		{title} {Bounding the sets of classical and quantum correlations in
			networks},\ }\href {https://doi.org/10.1103/PhysRevLett.123.140503}
	{\bibfield  {journal} {\bibinfo  {journal} {Phys. Rev. Lett.}\ }\textbf
		{\bibinfo {volume} {123}},\ \bibinfo {pages} {140503} (\bibinfo {year}
		{2019})}\BibitemShut {NoStop}%
	\bibitem [{\citenamefont {Bowles}\ \emph {et~al.}(2015)\citenamefont {Bowles},
		\citenamefont {Brunner},\ and\ \citenamefont
		{Paw\l{}owski}}]{PhysRevA.92.022351}%
	\BibitemOpen
	\bibfield  {author} {\bibinfo {author} {\bibfnamefont {J.}~\bibnamefont
			{Bowles}}, \bibinfo {author} {\bibfnamefont {N.}~\bibnamefont {Brunner}},\
		and\ \bibinfo {author} {\bibfnamefont {M.}~\bibnamefont {Paw\l{}owski}},\
	}\bibfield  {title} {\bibinfo {title} {Testing dimension and nonclassicality
			in communication networks},\ }\href
	{https://doi.org/10.1103/PhysRevA.92.022351} {\bibfield  {journal} {\bibinfo
			{journal} {Phys. Rev. A}\ }\textbf {\bibinfo {volume} {92}},\ \bibinfo
		{pages} {022351} (\bibinfo {year} {2015})}\BibitemShut {NoStop}%
	\bibitem [{\citenamefont {Wang}\ \emph {et~al.}(2019)\citenamefont {Wang},
		\citenamefont {Primaatmaja}, \citenamefont {Lavie}, \citenamefont
		{Varvitsiotis},\ and\ \citenamefont {Lim}}]{Wang2019}%
	\BibitemOpen
	\bibfield  {author} {\bibinfo {author} {\bibfnamefont {Y.}~\bibnamefont
			{Wang}}, \bibinfo {author} {\bibfnamefont {I.~W.}\ \bibnamefont
			{Primaatmaja}}, \bibinfo {author} {\bibfnamefont {E.}~\bibnamefont {Lavie}},
		\bibinfo {author} {\bibfnamefont {A.}~\bibnamefont {Varvitsiotis}},\ and\
		\bibinfo {author} {\bibfnamefont {C.~C.~W.}\ \bibnamefont {Lim}},\ }\bibfield
	{title} {\bibinfo {title} {Characterising the correlations of
			prepare-and-measure quantum networks},\ }\href
	{https://doi.org/10.1038/s41534-019-0133-3} {\bibfield  {journal} {\bibinfo
			{journal} {npj Quantum Information}\ }\textbf {\bibinfo {volume} {5}},\
		\bibinfo {pages} {17} (\bibinfo {year} {2019})}\BibitemShut {NoStop}%
	\bibitem [{\citenamefont {Silva}\ \emph {et~al.}(2015)\citenamefont {Silva},
		\citenamefont {Gisin}, \citenamefont {Guryanova},\ and\ \citenamefont
		{Popescu}}]{Silva_2015}%
	\BibitemOpen
	\bibfield  {author} {\bibinfo {author} {\bibfnamefont {R.}~\bibnamefont
			{Silva}}, \bibinfo {author} {\bibfnamefont {N.}~\bibnamefont {Gisin}},
		\bibinfo {author} {\bibfnamefont {Y.}~\bibnamefont {Guryanova}},\ and\
		\bibinfo {author} {\bibfnamefont {S.}~\bibnamefont {Popescu}},\ }\bibfield
	{title} {\bibinfo {title} {Multiple observers can share the nonlocality of
			half of an entangled pair by using optimal weak measurements},\ }\href
	{https://doi.org/10.1103/PhysRevLett.114.250401} {\bibfield  {journal}
		{\bibinfo  {journal} {Phys. Rev. Lett.}\ }\textbf {\bibinfo {volume} {114}},\
		\bibinfo {pages} {250401} (\bibinfo {year} {2015})}\BibitemShut {NoStop}%
	\bibitem [{\citenamefont {Curchod}\ \emph {et~al.}(2017)\citenamefont
		{Curchod}, \citenamefont {Johansson}, \citenamefont {Augusiak}, \citenamefont
		{Hoban}, \citenamefont {Wittek},\ and\ \citenamefont
		{Ac\'{\i}n}}]{PhysRevA.95.020102}%
	\BibitemOpen
	\bibfield  {author} {\bibinfo {author} {\bibfnamefont {F.~J.}\ \bibnamefont
			{Curchod}}, \bibinfo {author} {\bibfnamefont {M.}~\bibnamefont {Johansson}},
		\bibinfo {author} {\bibfnamefont {R.}~\bibnamefont {Augusiak}}, \bibinfo
		{author} {\bibfnamefont {M.~J.}\ \bibnamefont {Hoban}}, \bibinfo {author}
		{\bibfnamefont {P.}~\bibnamefont {Wittek}},\ and\ \bibinfo {author}
		{\bibfnamefont {A.}~\bibnamefont {Ac\'{\i}n}},\ }\bibfield  {title} {\bibinfo
		{title} {Unbounded randomness certification using sequences of
			measurements},\ }\href {https://doi.org/10.1103/PhysRevA.95.020102}
	{\bibfield  {journal} {\bibinfo  {journal} {Phys. Rev. A}\ }\textbf {\bibinfo
			{volume} {95}},\ \bibinfo {pages} {020102} (\bibinfo {year}
		{2017})}\BibitemShut {NoStop}%
	\bibitem [{\citenamefont {Liu}\ \emph {et~al.}(2024)\citenamefont {Liu},
		\citenamefont {Wang}, \citenamefont {Han},\ and\ \citenamefont
		{Wu}}]{Liu_2024}%
	\BibitemOpen
	\bibfield  {author} {\bibinfo {author} {\bibfnamefont {X.}~\bibnamefont
			{Liu}}, \bibinfo {author} {\bibfnamefont {Y.}~\bibnamefont {Wang}}, \bibinfo
		{author} {\bibfnamefont {Y.}~\bibnamefont {Han}},\ and\ \bibinfo {author}
		{\bibfnamefont {X.}~\bibnamefont {Wu}},\ }\bibfield  {title} {\bibinfo
		{title} {Quantifying the intrinsic randomness in sequential measurements},\
	}\href {https://doi.org/10.1088/1367-2630/ad19fe} {\bibfield  {journal}
		{\bibinfo  {journal} {New Journal of Physics}\ }\textbf {\bibinfo {volume}
			{26}},\ \bibinfo {pages} {013026} (\bibinfo {year} {2024})}\BibitemShut
	{NoStop}%
	\bibitem [{\citenamefont {Brown}\ and\ \citenamefont
		{Colbeck}(2020)}]{PhysRevLett.125.090401}%
	\BibitemOpen
	\bibfield  {author} {\bibinfo {author} {\bibfnamefont {P.~J.}\ \bibnamefont
			{Brown}}\ and\ \bibinfo {author} {\bibfnamefont {R.}~\bibnamefont
			{Colbeck}},\ }\bibfield  {title} {\bibinfo {title} {Arbitrarily many
			independent observers can share the nonlocality of a single maximally
			entangled qubit pair},\ }\href
	{https://doi.org/10.1103/PhysRevLett.125.090401} {\bibfield  {journal}
		{\bibinfo  {journal} {Phys. Rev. Lett.}\ }\textbf {\bibinfo {volume} {125}},\
		\bibinfo {pages} {090401} (\bibinfo {year} {2020})}\BibitemShut {NoStop}%
	\bibitem [{\citenamefont {Hu}\ \emph {et~al.}(2018)\citenamefont {Hu},
		\citenamefont {Zhou}, \citenamefont {Hu}, \citenamefont {Li}, \citenamefont
		{Guo},\ and\ \citenamefont {Zhang}}]{Hu2018}%
	\BibitemOpen
	\bibfield  {author} {\bibinfo {author} {\bibfnamefont {M.-J.}\ \bibnamefont
			{Hu}}, \bibinfo {author} {\bibfnamefont {Z.-Y.}\ \bibnamefont {Zhou}},
		\bibinfo {author} {\bibfnamefont {X.-M.}\ \bibnamefont {Hu}}, \bibinfo
		{author} {\bibfnamefont {C.-F.}\ \bibnamefont {Li}}, \bibinfo {author}
		{\bibfnamefont {G.-C.}\ \bibnamefont {Guo}},\ and\ \bibinfo {author}
		{\bibfnamefont {Y.-S.}\ \bibnamefont {Zhang}},\ }\bibfield  {title} {\bibinfo
		{title} {Observation of non-locality sharing among three observers with one
			entangled pair via optimal weak measurement},\ }\href
	{https://doi.org/10.1038/s41534-018-0115-x} {\bibfield  {journal} {\bibinfo
			{journal} {npj Quantum Information}\ }\textbf {\bibinfo {volume} {4}},\
		\bibinfo {pages} {63} (\bibinfo {year} {2018})}\BibitemShut {NoStop}%
	\bibitem [{\citenamefont {Sasmal}\ \emph {et~al.}(2018)\citenamefont {Sasmal},
		\citenamefont {Das}, \citenamefont {Mal},\ and\ \citenamefont
		{Majumdar}}]{PhysRevA.98.012305}%
	\BibitemOpen
	\bibfield  {author} {\bibinfo {author} {\bibfnamefont {S.}~\bibnamefont
			{Sasmal}}, \bibinfo {author} {\bibfnamefont {D.}~\bibnamefont {Das}},
		\bibinfo {author} {\bibfnamefont {S.}~\bibnamefont {Mal}},\ and\ \bibinfo
		{author} {\bibfnamefont {A.~S.}\ \bibnamefont {Majumdar}},\ }\bibfield
	{title} {\bibinfo {title} {Steering a single system sequentially by multiple
			observers},\ }\href {https://doi.org/10.1103/PhysRevA.98.012305} {\bibfield
		{journal} {\bibinfo  {journal} {Phys. Rev. A}\ }\textbf {\bibinfo {volume}
			{98}},\ \bibinfo {pages} {012305} (\bibinfo {year} {2018})}\BibitemShut
	{NoStop}%
	\bibitem [{\citenamefont {Bera}\ \emph {et~al.}(2018)\citenamefont {Bera},
		\citenamefont {Mal}, \citenamefont {Sen(De)},\ and\ \citenamefont
		{Sen}}]{PhysRevA.98.062304}%
	\BibitemOpen
	\bibfield  {author} {\bibinfo {author} {\bibfnamefont {A.}~\bibnamefont
			{Bera}}, \bibinfo {author} {\bibfnamefont {S.}~\bibnamefont {Mal}}, \bibinfo
		{author} {\bibfnamefont {A.}~\bibnamefont {Sen(De)}},\ and\ \bibinfo {author}
		{\bibfnamefont {U.}~\bibnamefont {Sen}},\ }\bibfield  {title} {\bibinfo
		{title} {Witnessing bipartite entanglement sequentially by multiple
			observers},\ }\href {https://doi.org/10.1103/PhysRevA.98.062304} {\bibfield
		{journal} {\bibinfo  {journal} {Phys. Rev. A}\ }\textbf {\bibinfo {volume}
			{98}},\ \bibinfo {pages} {062304} (\bibinfo {year} {2018})}\BibitemShut
	{NoStop}%
	\bibitem [{\citenamefont {Datta}\ and\ \citenamefont
		{Majumdar}(2018)}]{PhysRevA.98.042311}%
	\BibitemOpen
	\bibfield  {author} {\bibinfo {author} {\bibfnamefont {S.}~\bibnamefont
			{Datta}}\ and\ \bibinfo {author} {\bibfnamefont {A.~S.}\ \bibnamefont
			{Majumdar}},\ }\bibfield  {title} {\bibinfo {title} {Sharing of nonlocal
			advantage of quantum coherence by sequential observers},\ }\href
	{https://doi.org/10.1103/PhysRevA.98.042311} {\bibfield  {journal} {\bibinfo
			{journal} {Phys. Rev. A}\ }\textbf {\bibinfo {volume} {98}},\ \bibinfo
		{pages} {042311} (\bibinfo {year} {2018})}\BibitemShut {NoStop}%
	\bibitem [{\citenamefont {Bowles}\ \emph {et~al.}(2020)\citenamefont {Bowles},
		\citenamefont {Baccari},\ and\ \citenamefont
		{Salavrakos}}]{Bowles2020boundingsetsof}%
	\BibitemOpen
	\bibfield  {author} {\bibinfo {author} {\bibfnamefont {J.}~\bibnamefont
			{Bowles}}, \bibinfo {author} {\bibfnamefont {F.}~\bibnamefont {Baccari}},\
		and\ \bibinfo {author} {\bibfnamefont {A.}~\bibnamefont {Salavrakos}},\
	}\bibfield  {title} {\bibinfo {title} {Bounding sets of sequential quantum
			correlations and device-independent randomness certification},\ }\href
	{https://doi.org/10.22331/q-2020-10-19-344} {\bibfield  {journal} {\bibinfo
			{journal} {{Quantum}}\ }\textbf {\bibinfo {volume} {4}},\ \bibinfo {pages}
		{344} (\bibinfo {year} {2020})}\BibitemShut {NoStop}%
	\bibitem [{\citenamefont {Mohan}\ \emph {et~al.}(2019)\citenamefont {Mohan},
		\citenamefont {Tavakoli},\ and\ \citenamefont {Brunner}}]{Mohan_2019}%
	\BibitemOpen
	\bibfield  {author} {\bibinfo {author} {\bibfnamefont {K.}~\bibnamefont
			{Mohan}}, \bibinfo {author} {\bibfnamefont {A.}~\bibnamefont {Tavakoli}},\
		and\ \bibinfo {author} {\bibfnamefont {N.}~\bibnamefont {Brunner}},\
	}\bibfield  {title} {\bibinfo {title} {Sequential random access codes and
			self-testing of quantum measurement instruments},\ }\href
	{https://doi.org/10.1088/1367-2630/ab3773} {\bibfield  {journal} {\bibinfo
			{journal} {New J. Phys.}\ }\textbf {\bibinfo {volume} {21}},\ \bibinfo
		{pages} {083034} (\bibinfo {year} {2019})}\BibitemShut {NoStop}%
	\bibitem [{\citenamefont {Gerhardt}\ \emph {et~al.}(2011)\citenamefont
		{Gerhardt}, \citenamefont {Liu}, \citenamefont {Lamas-Linares}, \citenamefont
		{Skaar}, \citenamefont {Kurtsiefer},\ and\ \citenamefont
		{Makarov}}]{Gerhardt2011}%
	\BibitemOpen
	\bibfield  {author} {\bibinfo {author} {\bibfnamefont {I.}~\bibnamefont
			{Gerhardt}}, \bibinfo {author} {\bibfnamefont {Q.}~\bibnamefont {Liu}},
		\bibinfo {author} {\bibfnamefont {A.}~\bibnamefont {Lamas-Linares}}, \bibinfo
		{author} {\bibfnamefont {J.}~\bibnamefont {Skaar}}, \bibinfo {author}
		{\bibfnamefont {C.}~\bibnamefont {Kurtsiefer}},\ and\ \bibinfo {author}
		{\bibfnamefont {V.}~\bibnamefont {Makarov}},\ }\bibfield  {title} {\bibinfo
		{title} {Full-field implementation of a perfect eavesdropper on a quantum
			cryptography system},\ }\href {https://doi.org/10.1038/ncomms1348} {\bibfield
		{journal} {\bibinfo  {journal} {Nature Communications}\ }\textbf {\bibinfo
			{volume} {2}},\ \bibinfo {pages} {349} (\bibinfo {year} {2011})}\BibitemShut
	{NoStop}%
	\bibitem [{\citenamefont {Li}\ \emph {et~al.}(2021)\citenamefont {Li},
		\citenamefont {Zapatero}, \citenamefont {Tan}, \citenamefont {Wei},
		\citenamefont {Min}, \citenamefont {Liu}, \citenamefont {Jiang},
		\citenamefont {Liao}, \citenamefont {Peng}, \citenamefont {Curty},
		\citenamefont {Xu},\ and\ \citenamefont {Pan}}]{PhysRevApplied.15.034081}%
	\BibitemOpen
	\bibfield  {author} {\bibinfo {author} {\bibfnamefont {W.}~\bibnamefont
			{Li}}, \bibinfo {author} {\bibfnamefont {V.}~\bibnamefont {Zapatero}},
		\bibinfo {author} {\bibfnamefont {H.}~\bibnamefont {Tan}}, \bibinfo {author}
		{\bibfnamefont {K.}~\bibnamefont {Wei}}, \bibinfo {author} {\bibfnamefont
			{H.}~\bibnamefont {Min}}, \bibinfo {author} {\bibfnamefont {W.-Y.}\
			\bibnamefont {Liu}}, \bibinfo {author} {\bibfnamefont {X.}~\bibnamefont
			{Jiang}}, \bibinfo {author} {\bibfnamefont {S.-K.}\ \bibnamefont {Liao}},
		\bibinfo {author} {\bibfnamefont {C.-Z.}\ \bibnamefont {Peng}}, \bibinfo
		{author} {\bibfnamefont {M.}~\bibnamefont {Curty}}, \bibinfo {author}
		{\bibfnamefont {F.}~\bibnamefont {Xu}},\ and\ \bibinfo {author}
		{\bibfnamefont {J.-W.}\ \bibnamefont {Pan}},\ }\bibfield  {title} {\bibinfo
		{title} {Experimental quantum key distribution secure against malicious
			devices},\ }\href {https://doi.org/10.1103/PhysRevApplied.15.034081}
	{\bibfield  {journal} {\bibinfo  {journal} {Phys. Rev. Appl.}\ }\textbf
		{\bibinfo {volume} {15}},\ \bibinfo {pages} {034081} (\bibinfo {year}
		{2021})}\BibitemShut {NoStop}%
	\bibitem [{\citenamefont {Wath}\ \emph {et~al.}(2023)\citenamefont {Wath},
		\citenamefont {Hariprasad}, \citenamefont {Shah},\ and\ \citenamefont
		{Gupta}}]{Wath2023}%
	\BibitemOpen
	\bibfield  {author} {\bibinfo {author} {\bibfnamefont {Y.}~\bibnamefont
			{Wath}}, \bibinfo {author} {\bibfnamefont {M.}~\bibnamefont {Hariprasad}},
		\bibinfo {author} {\bibfnamefont {F.}~\bibnamefont {Shah}},\ and\ \bibinfo
		{author} {\bibfnamefont {S.}~\bibnamefont {Gupta}},\ }\bibfield  {title}
	{\bibinfo {title} {Eavesdropping a quantum key distribution network using
			sequential quantum unsharp measurement attacks},\ }\href
	{https://doi.org/10.1140/epjp/s13360-023-03664-4} {\bibfield  {journal}
		{\bibinfo  {journal} {The European Physical Journal Plus}\ }\textbf {\bibinfo
			{volume} {138}},\ \bibinfo {pages} {54} (\bibinfo {year} {2023})}\BibitemShut
	{NoStop}%
	\bibitem [{\citenamefont {Primaatmaja}\ \emph {et~al.}(2019)\citenamefont
		{Primaatmaja}, \citenamefont {Lavie}, \citenamefont {Goh}, \citenamefont
		{Wang},\ and\ \citenamefont {Lim}}]{PhysRevA.99.062332}%
	\BibitemOpen
	\bibfield  {author} {\bibinfo {author} {\bibfnamefont {I.~W.}\ \bibnamefont
			{Primaatmaja}}, \bibinfo {author} {\bibfnamefont {E.}~\bibnamefont {Lavie}},
		\bibinfo {author} {\bibfnamefont {K.~T.}\ \bibnamefont {Goh}}, \bibinfo
		{author} {\bibfnamefont {C.}~\bibnamefont {Wang}},\ and\ \bibinfo {author}
		{\bibfnamefont {C.~C.~W.}\ \bibnamefont {Lim}},\ }\bibfield  {title}
	{\bibinfo {title} {Versatile security analysis of
			measurement-device-independent quantum key distribution},\ }\href
	{https://doi.org/10.1103/PhysRevA.99.062332} {\bibfield  {journal} {\bibinfo
			{journal} {Phys. Rev. A}\ }\textbf {\bibinfo {volume} {99}},\ \bibinfo
		{pages} {062332} (\bibinfo {year} {2019})}\BibitemShut {NoStop}%
	\bibitem [{\citenamefont {Pereira}\ \emph {et~al.}(2019)\citenamefont
		{Pereira}, \citenamefont {Curty},\ and\ \citenamefont
		{Tamaki}}]{Pereira2019}%
	\BibitemOpen
	\bibfield  {author} {\bibinfo {author} {\bibfnamefont {M.}~\bibnamefont
			{Pereira}}, \bibinfo {author} {\bibfnamefont {M.}~\bibnamefont {Curty}},\
		and\ \bibinfo {author} {\bibfnamefont {K.}~\bibnamefont {Tamaki}},\
	}\bibfield  {title} {\bibinfo {title} {Quantum key distribution with flawed
			and leaky sources},\ }\href {https://doi.org/10.1038/s41534-019-0180-9}
	{\bibfield  {journal} {\bibinfo  {journal} {npj Quantum Information}\
		}\textbf {\bibinfo {volume} {5}},\ \bibinfo {pages} {62} (\bibinfo {year}
		{2019})}\BibitemShut {NoStop}%
	\bibitem [{\citenamefont {Liu}\ \emph {et~al.}(2019)\citenamefont {Liu},
		\citenamefont {Wang}, \citenamefont {Lavie}, \citenamefont {Wang},
		\citenamefont {Ricou}, \citenamefont {Guo},\ and\ \citenamefont
		{Lim}}]{PhysRevApplied.12.024048}%
	\BibitemOpen
	\bibfield  {author} {\bibinfo {author} {\bibfnamefont {L.}~\bibnamefont
			{Liu}}, \bibinfo {author} {\bibfnamefont {Y.}~\bibnamefont {Wang}}, \bibinfo
		{author} {\bibfnamefont {E.}~\bibnamefont {Lavie}}, \bibinfo {author}
		{\bibfnamefont {C.}~\bibnamefont {Wang}}, \bibinfo {author} {\bibfnamefont
			{A.}~\bibnamefont {Ricou}}, \bibinfo {author} {\bibfnamefont {F.~Z.}\
			\bibnamefont {Guo}},\ and\ \bibinfo {author} {\bibfnamefont {C.~C.~W.}\
			\bibnamefont {Lim}},\ }\bibfield  {title} {\bibinfo {title} {Practical
			quantum key distribution with non-phase-randomized coherent states},\ }\href
	{https://doi.org/10.1103/PhysRevApplied.12.024048} {\bibfield  {journal}
		{\bibinfo  {journal} {Phys. Rev. Appl.}\ }\textbf {\bibinfo {volume} {12}},\
		\bibinfo {pages} {024048} (\bibinfo {year} {2019})}\BibitemShut {NoStop}%
	\bibitem [{\citenamefont {Li}\ \emph {et~al.}(2012)\citenamefont {Li},
		\citenamefont {Paw\l{}owski}, \citenamefont {Yin}, \citenamefont {Guo},\ and\
		\citenamefont {Han}}]{PhysRevA.85.052308}%
	\BibitemOpen
	\bibfield  {author} {\bibinfo {author} {\bibfnamefont {H.-W.}\ \bibnamefont
			{Li}}, \bibinfo {author} {\bibfnamefont {M.}~\bibnamefont {Paw\l{}owski}},
		\bibinfo {author} {\bibfnamefont {Z.-Q.}\ \bibnamefont {Yin}}, \bibinfo
		{author} {\bibfnamefont {G.-C.}\ \bibnamefont {Guo}},\ and\ \bibinfo {author}
		{\bibfnamefont {Z.-F.}\ \bibnamefont {Han}},\ }\bibfield  {title} {\bibinfo
		{title} {Semi-device-independent randomness certification using
			$n\ensuremath{\rightarrow}1$ quantum random access codes},\ }\href
	{https://doi.org/10.1103/PhysRevA.85.052308} {\bibfield  {journal} {\bibinfo
			{journal} {Phys. Rev. A}\ }\textbf {\bibinfo {volume} {85}},\ \bibinfo
		{pages} {052308} (\bibinfo {year} {2012})}\BibitemShut {NoStop}%
	\bibitem [{\citenamefont {Wilde}(2017)}]{Wilde_2017}%
	\BibitemOpen
	\bibfield  {author} {\bibinfo {author} {\bibfnamefont {M.~M.}\ \bibnamefont
			{Wilde}},\ }\href@noop {} {\emph {\bibinfo {title} {Quantum Information
				Theory}}},\ \bibinfo {edition} {2nd}\ ed.\ (\bibinfo  {publisher} {Cambridge
		University Press},\ \bibinfo {year} {2017})\BibitemShut {NoStop}%
	\bibitem [{\citenamefont {Tavakoli}\ \emph {et~al.}(2015)\citenamefont
		{Tavakoli}, \citenamefont {Hameedi}, \citenamefont {Marques},\ and\
		\citenamefont {Bourennane}}]{PhysRevLett.114.170502}%
	\BibitemOpen
	\bibfield  {author} {\bibinfo {author} {\bibfnamefont {A.}~\bibnamefont
			{Tavakoli}}, \bibinfo {author} {\bibfnamefont {A.}~\bibnamefont {Hameedi}},
		\bibinfo {author} {\bibfnamefont {B.}~\bibnamefont {Marques}},\ and\ \bibinfo
		{author} {\bibfnamefont {M.}~\bibnamefont {Bourennane}},\ }\bibfield  {title}
	{\bibinfo {title} {Quantum random access codes using single $d$-level
			systems},\ }\href {https://doi.org/10.1103/PhysRevLett.114.170502} {\bibfield
		{journal} {\bibinfo  {journal} {Phys. Rev. Lett.}\ }\textbf {\bibinfo
			{volume} {114}},\ \bibinfo {pages} {170502} (\bibinfo {year}
		{2015})}\BibitemShut {NoStop}%
	\bibitem [{\citenamefont {Tavakoli}\ \emph {et~al.}(2018)\citenamefont
		{Tavakoli}, \citenamefont {Kaniewski}, \citenamefont {V\'ertesi},
		\citenamefont {Rosset},\ and\ \citenamefont {Brunner}}]{PhysRevA.98.062307}%
	\BibitemOpen
	\bibfield  {author} {\bibinfo {author} {\bibfnamefont {A.}~\bibnamefont
			{Tavakoli}}, \bibinfo {author} {\bibfnamefont {J.~m.~k.}\ \bibnamefont
			{Kaniewski}}, \bibinfo {author} {\bibfnamefont {T.}~\bibnamefont
			{V\'ertesi}}, \bibinfo {author} {\bibfnamefont {D.}~\bibnamefont {Rosset}},\
		and\ \bibinfo {author} {\bibfnamefont {N.}~\bibnamefont {Brunner}},\
	}\bibfield  {title} {\bibinfo {title} {Self-testing quantum states and
			measurements in the prepare-and-measure scenario},\ }\href
	{https://doi.org/10.1103/PhysRevA.98.062307} {\bibfield  {journal} {\bibinfo
			{journal} {Phys. Rev. A}\ }\textbf {\bibinfo {volume} {98}},\ \bibinfo
		{pages} {062307} (\bibinfo {year} {2018})}\BibitemShut {NoStop}%
	\bibitem [{\citenamefont {Li}\ \emph {et~al.}(2011)\citenamefont {Li},
		\citenamefont {Yin}, \citenamefont {Wu}, \citenamefont {Zou}, \citenamefont
		{Wang}, \citenamefont {Chen}, \citenamefont {Guo},\ and\ \citenamefont
		{Han}}]{PhysRevA.84.034301}%
	\BibitemOpen
	\bibfield  {author} {\bibinfo {author} {\bibfnamefont {H.-W.}\ \bibnamefont
			{Li}}, \bibinfo {author} {\bibfnamefont {Z.-Q.}\ \bibnamefont {Yin}},
		\bibinfo {author} {\bibfnamefont {Y.-C.}\ \bibnamefont {Wu}}, \bibinfo
		{author} {\bibfnamefont {X.-B.}\ \bibnamefont {Zou}}, \bibinfo {author}
		{\bibfnamefont {S.}~\bibnamefont {Wang}}, \bibinfo {author} {\bibfnamefont
			{W.}~\bibnamefont {Chen}}, \bibinfo {author} {\bibfnamefont {G.-C.}\
			\bibnamefont {Guo}},\ and\ \bibinfo {author} {\bibfnamefont {Z.-F.}\
			\bibnamefont {Han}},\ }\bibfield  {title} {\bibinfo {title}
		{Semi-device-independent random-number expansion without entanglement},\
	}\href {https://doi.org/10.1103/PhysRevA.84.034301} {\bibfield  {journal}
		{\bibinfo  {journal} {Phys. Rev. A}\ }\textbf {\bibinfo {volume} {84}},\
		\bibinfo {pages} {034301} (\bibinfo {year} {2011})}\BibitemShut {NoStop}%
	\bibitem [{\citenamefont {Zhou}\ \emph {et~al.}(2015)\citenamefont {Zhou},
		\citenamefont {Li}, \citenamefont {Wang}, \citenamefont {Li}, \citenamefont
		{Gao},\ and\ \citenamefont {Wen}}]{PhysRevA.92.022331}%
	\BibitemOpen
	\bibfield  {author} {\bibinfo {author} {\bibfnamefont {Y.-Q.}\ \bibnamefont
			{Zhou}}, \bibinfo {author} {\bibfnamefont {H.-W.}\ \bibnamefont {Li}},
		\bibinfo {author} {\bibfnamefont {Y.-K.}\ \bibnamefont {Wang}}, \bibinfo
		{author} {\bibfnamefont {D.-D.}\ \bibnamefont {Li}}, \bibinfo {author}
		{\bibfnamefont {F.}~\bibnamefont {Gao}},\ and\ \bibinfo {author}
		{\bibfnamefont {Q.-Y.}\ \bibnamefont {Wen}},\ }\bibfield  {title} {\bibinfo
		{title} {Semi-device-independent randomness expansion with partially free
			random sources},\ }\href {https://doi.org/10.1103/PhysRevA.92.022331}
	{\bibfield  {journal} {\bibinfo  {journal} {Phys. Rev. A}\ }\textbf {\bibinfo
			{volume} {92}},\ \bibinfo {pages} {022331} (\bibinfo {year}
		{2015})}\BibitemShut {NoStop}%
	\bibitem [{\citenamefont {Passaro}\ \emph {et~al.}(2015)\citenamefont
		{Passaro}, \citenamefont {Cavalcanti}, \citenamefont {Skrzypczyk},\ and\
		\citenamefont {Acín}}]{Passaro_2015}%
	\BibitemOpen
	\bibfield  {author} {\bibinfo {author} {\bibfnamefont {E.}~\bibnamefont
			{Passaro}}, \bibinfo {author} {\bibfnamefont {D.}~\bibnamefont {Cavalcanti}},
		\bibinfo {author} {\bibfnamefont {P.}~\bibnamefont {Skrzypczyk}},\ and\
		\bibinfo {author} {\bibfnamefont {A.}~\bibnamefont {Acín}},\ }\bibfield
	{title} {\bibinfo {title} {Optimal randomness certification in the quantum
			steering and prepare-and-measure scenarios},\ }\href
	{https://doi.org/10.1088/1367-2630/17/11/113010} {\bibfield  {journal}
		{\bibinfo  {journal} {New J. Phys.}\ }\textbf {\bibinfo {volume} {17}},\
		\bibinfo {pages} {113010} (\bibinfo {year} {2015})}\BibitemShut {NoStop}%
	\bibitem [{\citenamefont {Brask}\ \emph {et~al.}(2017)\citenamefont {Brask},
		\citenamefont {Martin}, \citenamefont {Esposito}, \citenamefont {Houlmann},
		\citenamefont {Bowles}, \citenamefont {Zbinden},\ and\ \citenamefont
		{Brunner}}]{PhysRevApplied.7.054018}%
	\BibitemOpen
	\bibfield  {author} {\bibinfo {author} {\bibfnamefont {J.~B.}\ \bibnamefont
			{Brask}}, \bibinfo {author} {\bibfnamefont {A.}~\bibnamefont {Martin}},
		\bibinfo {author} {\bibfnamefont {W.}~\bibnamefont {Esposito}}, \bibinfo
		{author} {\bibfnamefont {R.}~\bibnamefont {Houlmann}}, \bibinfo {author}
		{\bibfnamefont {J.}~\bibnamefont {Bowles}}, \bibinfo {author} {\bibfnamefont
			{H.}~\bibnamefont {Zbinden}},\ and\ \bibinfo {author} {\bibfnamefont
			{N.}~\bibnamefont {Brunner}},\ }\bibfield  {title} {\bibinfo {title}
		{Megahertz-rate semi-device-independent quantum random number generators
			based on unambiguous state discrimination},\ }\href
	{https://doi.org/10.1103/PhysRevApplied.7.054018} {\bibfield  {journal}
		{\bibinfo  {journal} {Phys. Rev. Appl.}\ }\textbf {\bibinfo {volume} {7}},\
		\bibinfo {pages} {054018} (\bibinfo {year} {2017})}\BibitemShut {NoStop}%
	\bibitem [{\citenamefont {Xiao}\ \emph {et~al.}(2023)\citenamefont {Xiao},
		\citenamefont {Guo}, \citenamefont {Dong},\ and\ \citenamefont
		{Gao}}]{Xiao2023}%
	\BibitemOpen
	\bibfield  {author} {\bibinfo {author} {\bibfnamefont {Y.}~\bibnamefont
			{Xiao}}, \bibinfo {author} {\bibfnamefont {F.}~\bibnamefont {Guo}}, \bibinfo
		{author} {\bibfnamefont {H.}~\bibnamefont {Dong}},\ and\ \bibinfo {author}
		{\bibfnamefont {F.}~\bibnamefont {Gao}},\ }\bibfield  {title} {\bibinfo
		{title} {Expanding the sharpness parameter area based on sequential
			{$3{\rightarrow }1$} parity-oblivious quantum random access code},\ }\href
	{https://doi.org/10.1007/s11128-023-03924-3} {\bibfield  {journal} {\bibinfo
			{journal} {Quantum Information Processing}\ }\textbf {\bibinfo {volume}
			{22}},\ \bibinfo {pages} {195} (\bibinfo {year} {2023})}\BibitemShut
	{NoStop}%
	\bibitem [{\citenamefont {Nieto-Silleras}\ \emph {et~al.}(2014)\citenamefont
		{Nieto-Silleras}, \citenamefont {Pironio},\ and\ \citenamefont
		{Silman}}]{Nieto-Silleras_2014}%
	\BibitemOpen
	\bibfield  {author} {\bibinfo {author} {\bibfnamefont {O.}~\bibnamefont
			{Nieto-Silleras}}, \bibinfo {author} {\bibfnamefont {S.}~\bibnamefont
			{Pironio}},\ and\ \bibinfo {author} {\bibfnamefont {J.}~\bibnamefont
			{Silman}},\ }\bibfield  {title} {\bibinfo {title} {Using complete measurement
			statistics for optimal device-independent randomness evaluation},\ }\href
	{https://doi.org/10.1088/1367-2630/16/1/013035} {\bibfield  {journal}
		{\bibinfo  {journal} {New Journal of Physics}\ }\textbf {\bibinfo {volume}
			{16}},\ \bibinfo {pages} {013035} (\bibinfo {year} {2014})}\BibitemShut
	{NoStop}%
	\bibitem [{\citenamefont {Bancal}\ \emph {et~al.}(2014)\citenamefont {Bancal},
		\citenamefont {Sheridan},\ and\ \citenamefont {Scarani}}]{Bancal_2014}%
	\BibitemOpen
	\bibfield  {author} {\bibinfo {author} {\bibfnamefont {J.-D.}\ \bibnamefont
			{Bancal}}, \bibinfo {author} {\bibfnamefont {L.}~\bibnamefont {Sheridan}},\
		and\ \bibinfo {author} {\bibfnamefont {V.}~\bibnamefont {Scarani}},\
	}\bibfield  {title} {\bibinfo {title} {More randomness from the same data},\
	}\href {https://doi.org/10.1088/1367-2630/16/3/033011} {\bibfield  {journal}
		{\bibinfo  {journal} {New Journal of Physics}\ }\textbf {\bibinfo {volume}
			{16}},\ \bibinfo {pages} {033011} (\bibinfo {year} {2014})}\BibitemShut
	{NoStop}%
	\bibitem [{\citenamefont {Canabarro}\ \emph {et~al.}(2019)\citenamefont
		{Canabarro}, \citenamefont {Brito},\ and\ \citenamefont
		{Chaves}}]{PhysRevLett.122.200401}%
	\BibitemOpen
	\bibfield  {author} {\bibinfo {author} {\bibfnamefont {A.}~\bibnamefont
			{Canabarro}}, \bibinfo {author} {\bibfnamefont {S.}~\bibnamefont {Brito}},\
		and\ \bibinfo {author} {\bibfnamefont {R.}~\bibnamefont {Chaves}},\
	}\bibfield  {title} {\bibinfo {title} {Machine learning nonlocal
			correlations},\ }\href {https://doi.org/10.1103/PhysRevLett.122.200401}
	{\bibfield  {journal} {\bibinfo  {journal} {Phys. Rev. Lett.}\ }\textbf
		{\bibinfo {volume} {122}},\ \bibinfo {pages} {200401} (\bibinfo {year}
		{2019})}\BibitemShut {NoStop}%
	\bibitem [{\citenamefont {Song}\ \emph {et~al.}(2024)\citenamefont {Song},
		\citenamefont {Wu}, \citenamefont {Wu}, \citenamefont {Li}, \citenamefont
		{Wen}, \citenamefont {Qin},\ and\ \citenamefont {Gao}}]{Song2024}%
	\BibitemOpen
	\bibfield  {author} {\bibinfo {author} {\bibfnamefont {Y.}~\bibnamefont
			{Song}}, \bibinfo {author} {\bibfnamefont {Y.}~\bibnamefont {Wu}}, \bibinfo
		{author} {\bibfnamefont {S.}~\bibnamefont {Wu}}, \bibinfo {author}
		{\bibfnamefont {D.}~\bibnamefont {Li}}, \bibinfo {author} {\bibfnamefont
			{Q.}~\bibnamefont {Wen}}, \bibinfo {author} {\bibfnamefont {S.}~\bibnamefont
			{Qin}},\ and\ \bibinfo {author} {\bibfnamefont {F.}~\bibnamefont {Gao}},\
	}\bibfield  {title} {\bibinfo {title} {A quantum federated learning framework
			for classical clients},\ }\href {https://doi.org/10.1007/s11433-023-2337-2}
	{\bibfield  {journal} {\bibinfo  {journal} {Science China Physics, Mechanics
				\& Astronomy}\ }\textbf {\bibinfo {volume} {67}},\ \bibinfo {pages} {250311}
		(\bibinfo {year} {2024})}\BibitemShut {NoStop}%
	\bibitem [{\citenamefont {D'Alessandro}\ \emph {et~al.}(2023)\citenamefont
		{D'Alessandro}, \citenamefont {Polacchi}, \citenamefont {Moreno},
		\citenamefont {Polino}, \citenamefont {Chaves}, \citenamefont {Agresti},\
		and\ \citenamefont {Sciarrino}}]{PhysRevResearch.5.023016}%
	\BibitemOpen
	\bibfield  {author} {\bibinfo {author} {\bibfnamefont {N.}~\bibnamefont
			{D'Alessandro}}, \bibinfo {author} {\bibfnamefont {B.}~\bibnamefont
			{Polacchi}}, \bibinfo {author} {\bibfnamefont {G.}~\bibnamefont {Moreno}},
		\bibinfo {author} {\bibfnamefont {E.}~\bibnamefont {Polino}}, \bibinfo
		{author} {\bibfnamefont {R.}~\bibnamefont {Chaves}}, \bibinfo {author}
		{\bibfnamefont {I.}~\bibnamefont {Agresti}},\ and\ \bibinfo {author}
		{\bibfnamefont {F.}~\bibnamefont {Sciarrino}},\ }\bibfield  {title} {\bibinfo
		{title} {Machine-learning-based device-independent certification of quantum
			networks},\ }\href {https://doi.org/10.1103/PhysRevResearch.5.023016}
	{\bibfield  {journal} {\bibinfo  {journal} {Phys. Rev. Res.}\ }\textbf
		{\bibinfo {volume} {5}},\ \bibinfo {pages} {023016} (\bibinfo {year}
		{2023})}\BibitemShut {NoStop}%
\end{thebibliography}
%

\end{document}